

\voffset=-2pt \hoffset=-3.5pt \vsize=56pc \hsize=39pc
\baselineskip=18pt \parskip=3pt plus 2pt minus 1pt \tolerance=1000
\hfuzz=75pt
 
\font\fr=cmr10 at 10pt
\font\fvlbx=cmbx10 at 14.4pt \font\flbx=cmbx10 at 12pt
\font\fbx=cmbx10 at 10pt
 \font\fss=cmss10 at 10pt
\font\ftt=cmtt10 at 10pt
 \font\fit=cmti10 at 10pt
\fss 
\font\fsl=cmsl10 at 10pt  
\def\i#1#2{\item{\fbx #1}{\fr #2}}
\def\ii#1#2{\itemitem{\fbx #1}{\fr #2}}
\def\dsq{\buildrel {\hskip -1.2pt \bullet} \over \sq}
\raggedbottom
\def\dpar{\buildrel \bullet \over \partial}  \def\pr{\prime}
\def\sqr#1#2{{\vcenter{\hrule height.#2pt \hbox{\vrule width.#2pt
height#1pt \kern#1pt \vrule width.#2pt} \hrule height.#2pt}}}
\def\toda{\number\day \space\ifcase\month\or Jan\or Feb\or Mar\or
Apr\or
May\or Jun\or Jly\or Aug\or Sep\or Oct\or Nov\or Dec\fi
\space\number\year}
\def\sq{{\mathchoice\sqr55\sqr55\sqr{2.1}3\sqr{1.5}3}\hskip 1.5pt}
\def\lrhup#1{\buildrel {{\leftharpoonup \hskip -8pt
\rightharpoonup}} \over #1}
\def\sect#1{\vskip 10pt \ni{\flbx #1} \nobreak \vskip 3pt}
\def\cl{\centerline}

\def\ni{\noindent}
\def\insect#1{\vskip 5pt \ni{\fbx #1}\ \ }
\def\ent#1#2{{{\baselineskip=13pt \vskip 1pt \i{\fbx #1}{\fr{#2}
\vskip 9pt}}}}
 \def\perm#1{\hskip 2pt #1 {\rm - perms}}

\def\eqnn#1{\eqno{\hbox{(#1)}}} 

\def\dpr{{\prime\prime}}
 
 \def\ruline{\vskip 8pt \hrule
\vskip 8pt}

\null
\vskip 1pc
\footline={}

\cl{\fvlbx A Cutoff Procedure and Counterterms for}

\cl{\fvlbx Differential Renormalization}

\baselineskip=12pt  \vskip 2pc

\cl{{\bf Daniel Z. Freedman}\footnote{$^1$}{e-mail:
dzf@math.mit.edu.}}

\cl{\it Department of Mathematics}
\cl{and}
\cl{\it Center for Theoretical Physics}
\cl{\it Massachusetts Institute of Technology}
\cl{\it Cambridge, Massachusetts\ \ 02139\ \ \ U.S.A.}

\vskip 1pc

\cl{{\bf Kenneth Johnson}\footnote{$^2$}{e-mail:
knjhnsn@mitlns.bitnet}}

\cl{\it Center for Theoretical Physics}
\cl{\it Laboratory for Nuclear Science}
\cl{\it and Department of Physics}
\cl{\it Massachusetts Institute of Technology}
\cl{\it Cambridge, Massachusetts\ \ 02139\ \ \ U.S.A.}

\vskip 1pc

\cl{{\bf Ramon Mu\~noz-Tapia}\footnote{$^3$}{e-mail:
rmt@hep.dur.ac.uk}}

\cl{\it Department of Physics}
\cl{\it University of Durham}
\cl{\it Science Laboratories, South Road}
\cl{\it Durham, DH1 3LE, U.K.}

\vskip .5pc

\cl{and}

\vskip .5pc

\cl{{\bf Xavier Vilas{\' {\char'20}}s-Cardona}\footnote{$^4$}{e-mail:
druida@ebubecm1.bitnet}}

\cl{\it Departament d'Estructura}
\cl{\it i Constituents de la Mat\`eria}
\cl{\it Universitat de Barcelona}
\cl{\it Barcelona\ \ SPAIN}

\vskip 1pc

\cl{Typeset in \TeX\ by Martha Adams and Roger L. Gilson.}

\vfill

\ni{\bf CTP\# 2099}

\ni{\bf DTP/92/40}

\ni{{\bf UB-ECM-PF 92/13} \hfill {\bf May 1992}} \eject

\null  \baselineskip 16pt 
\vskip 1pc

\cl{\fvlbx A Cutoff Procedure and Counterterms for}

\cl{\fvlbx Differential Renormalization}

\vskip 1pc

\cl{\fr D.Z.~Freedman, K.~Johnson, R.~Mu\~noz-Tapia,
X.~Vilas{\'{\char'20}}s-Cardona}

\vskip 2pc

\cl{\flbx Abstract}

\vskip 1pc

Explicit divergences and counterterms do not appear in the
differential renormalization method, but they are concealed in the
neglected surface terms in the formal partial integration procedure
used.  A systematic real space cutoff procedure for massless $\phi^4$
theory is therefore studied in order to test the method and its
compatibility with unitarity.  Through 3-loop order, it is found that
cutoff bare amplitudes are equal to the renormalized amplitudes
previously obtained using the formal procedure plus singular terms
which can be consistently cancelled by adding conventional
counterterms to the Lagrangian.  Renormalization group functions
$\beta (g)$ and $\gamma (g)$ obtained in the cutoff theory also agree
with previous results.



\sect{1.\ \ Introduction}

\pageno=1
\footline={\hss \folio \hss} \baselineskip 14pt

The purpose of this paper is to explore another aspect of the recently
proposed differential renormalization procedure [1].  That method
relies on the observation that (essentially all) primitively divergent
Feynman graphs are well defined in real space for non-coincident
points, but too singular at short distance to allow a Fourier
transform.  A regularization procedure must supply a prescription for
the real space amplitudes which defines the short distance
singularities such that integrals over them are well defined.  This
was done in [1] by a method which simultaneously regularizes and
renormalizes amplitudes.  The ideas involved are quite simple and best
stated in terms of the 1-loop 4-point bubble graph of massless
$\phi^4$
theory in 4 dimensions.  This involves the singular function $1/x^4$,
where $x^4 = \left( x_\mu x_\mu \right)^2$, and is regulated as
follows:

\i{1.}{\fss Express such singular functions as derivatives of other
functions which have well defined Fourier transforms.  For example,
$$
{1 \over x^4} = - {1 \over 4} \sq {{\ln M^2 x^2} \over x^2} \eqnn{1.1}
$$
is an identity for $x \not = 0$, and the function $\ln \left( M^2 x^2
\right) / x^2$
has Fourier transform $-4 \pi^2 \ln \left( p^2 / \overline M^2
\right)
/p^2$ where $\overline M = 2M / \gamma$ and $\gamma = 1.781 \ldots$
is Euler's constant.}

\i{2.}{\fss Use formal partial integration
of the derivatives in (1.1) to compute integrals such as the
Fourier transform.  Thus the regulated Fourier transform of $1/x^4$ is
defined as $- \pi^2 \ln \left( p^2 / \overline M^2 \right)$.}

In Ref.~[1] it was shown in a very explicit study of massless $\phi^4$
theory through 3-loop order, that these ideas can be extended to
renormalize all 1PI vertex functions, including both primitively
divergent graphs and those with divergent subgraphs.  It was also
shown that the resulting amplitudes satisfy the renormalization group
equations in which $M$ appears as the expected scale variable.
Further applications of differential renormalization to gauge
theories, supersymmetry, and amplitudes with massive particles have
recently appeared [2].

Explicit divergences and the counterterms which cancel them never occur
in differential renormalization.  The usual ultraviolet divergences of
field theory are hidden in the short distance surface terms which are
dropped in step 2 above.  It was an implicit article-of-faith in [1],
justified only in 1-loop order, that these surface terms could be
cancelled by counterterms for wave function, mass, and coupling
renormalizations.  Since this is crucial to the consistency of the
procedure, we undertake to demonstrate it here up to 3-loop order in
$\phi^4$ theory.  For this purpose we wish to repeat the calculations
of [1] using an explicit cutoff, implemented by modifying the
Euclidean massless scalar propagator as follows,
$$
\Delta (x) = {1 \over {4 \pi^2}} {1 \over x^2} \ \longrightarrow \ \Delta
(x,\varepsilon) = {1 \over {4 \pi^2}} {1 \over {x^2 + \varepsilon^2}} \
. \eqnn{1.2}
$$
Real space calculations with this propagator are modelled as closely
as possible on the differential methods of [1], leading to bare
amplitudes $\Gamma^b (x_i, \varepsilon)$.  In the limit of small
$\varepsilon$, we show that the bare amplitudes for each diagram can
be expressed as the renormalized amplitudes of [1] $\Gamma^r (x,M)$
plus additional singular terms involving $1/\varepsilon^2$ or $\left(
\ln \varepsilon^2 M^2 \right)^n$.  The latter are cancelled by adding
local counterterms to the Lagrangian and including graphs generated
by counterterm vertices.  The scale $M$ is required for
dimensional reasons in the separation of regular and singular terms as
$\varepsilon \rightarrow 0$.  It will be clear from our calculations
that the singular terms are related to the surface terms neglected in
[1], and the consistency of Step 2 above is thereby demonstrated.

This investigation was primarily motivated by skeptics of the methods
of [1] who were not convinced that overlap divergences were treated
correctly and suspected an attendant violation of unitarity.  We
believe that the present investigation resolves such doubts at the
concrete level of current calculations.  Specifically, the fact that
the singular cutoff dependence of bare amplitudes is cancelled by a
Hermitean modification of the Lagrangian effectively proves that the
results of [1] are unitary, provided that any non-unitarity in the
cutoff chosen vanishes as the cutoff is removed.  In the present case
the Fourier transform of the cutoff propagator is
$$
\Delta (x,\varepsilon) = \int{{d^4 p} \over {(2 \pi)^4}} e^{ip \cdot
x} {\varepsilon \over p} K_1 (\varepsilon p) \eqnn{1.3}
$$
and contains a logarithmic branch point at $p = 0$, with the cut
corresponding to time-like Lorentzian momentum.  This behavior is
displayed in the limiting form for small $\varepsilon$ and fixed $p$,
$$
{\varepsilon \over p} K_1 (\varepsilon p) \sim {1 \over p^2} + {1
\over 4} \varepsilon^2 \left[ \ln \left( {{\varepsilon^2 p^2 \gamma^2}
\over 4} \right) - 1 \right] + {\cal O} (\varepsilon^4 ) \ . \eqnn{1.4}
$$
Unitarity is satisfied as $\varepsilon \rightarrow 0$ since effects of
the logarithm vanish quadratically and the quadratic divergences in
cutoff amplitudes are directly cancelled by additive mass counterterms,
so there are no $1 / \varepsilon^2$ terms which multiply (1.4).

One should note that ${\varepsilon \over p} K_1 (\varepsilon p)$ falls
exponentially as $p \rightarrow \infty$.
Our real space computations are therefore equivalent to a
momentum space approach with a damped propagator similar to that used
in [3] to study the renormalization group in quantum field
theory.

Renormalization group equations are usually derived [4] by
studying the relation between bare and renormalized amplitudes.  Since
we have now systematically defined cutoff bare amplitudes within the
differential renormalization method, we can repeat this derivation and
obtain $\beta (g)$ and $\gamma (g)$ as conventional scale derivatives
of the renormalization constants.  Our results agree with those found
using the ``experimental'' approach to the renormalization group
equations taken in [1].


\sect{2.\ \ The Cutoff Method in 1-Loop Order}

The Lagrangian of massive Euclidean signature $\phi^4$ theory is
$$
{\cal L} = {1 \over 2} (\partial \phi)^2 + {1 \over 2} m^2 \phi^2 + {1
\over {24}} \lambda \phi^4\ .  \eqnn{2.1}
$$
We will be concerned only with the massless case in this paper.  The
bare amplitudes for all Feynman diagrams will be computed using the
Feynman rules of (2.1) with the cutoff propagator (1.2).  These
Feynman diagrams are shown in Figs 1--2, in which the same designation
as in [1] is used.

\vbox{ \ruline {\fbx See figure 1 of Reference [1]}
\item{Fig.~1:}{\fit Diagrams which contribute to $\Gamma^{(2)}(x-y)$ in
$\phi^4$ theory.  The divergence associated to the tadpole diagram
{\ftt b}\ is immediately cancelled by an appropriate counterterm in
our renormalization scheme, so 2- and 3-loop graphs which include
tadpoles need not be considered.} \ruline }

The classical contribution to the 4-point function, diagram {\ftt e\/},
and the
bare amplitude for the 1-loop bubble graph {\ftt f\/}, are given by
$$
\Gamma_e (x_i) = - 16 \pi^2 g \delta_{12} \delta_{13} \delta_{14}
\eqnn{2.2}
$$ and $$
\Gamma_f^b (x_i , \varepsilon ) = 8 g^2 \delta_{12} \delta_{34} {1
\over {\left[ x_{13}^2 + \varepsilon^2 \right]^2}} + \perm{2}
\eqnn{2.3}
$$ where $$
g = {\lambda \over {16 \pi^2}} \ , \eqnn{2.4}
$$ and the notation $$ \eqalign{
x_{i j} &= x_i - x_j \cr
\delta_{ij} &= \delta^4 (x_i - x_j) \cr } \eqnn{2.5}
$$
will be used throughout.

As in Section I of [1], we express $1/(x^2 + \varepsilon^2)^2$ as $\sq
G (x^2 , \varepsilon^2 )$, leading to the ordinary linear differential
equation
$$
{1 \over {(z + \varepsilon^2)^2}} = {4 \over z} {d \over {dz}} \left(
z^2
\, {{dG} \over {dz}} \right) \eqnn{2.6}
$$
in the variable $z = x^2$.  The general solution is
$$
G(x^2 , \varepsilon^2 ) = - {1 \over 4} {{\ln [(x^2 + \varepsilon^2
)/\varepsilon^2]} \over x^2} + {a \over x^2} + b \ . \eqnn{2.7}
$$
The additive constant $b$ is irrelevant and can be dropped.  The basic
differential identity then becomes
$$
{1 \over {(x^2 + \varepsilon^2)^2}} = - {1 \over 4} \sq {{\ln \left[
\left( x^2 + \varepsilon^2 \right) / \varepsilon^2 \right]} \over x^2}
\ . \eqnn{2.8}
$$
We have chosen $a=0$ because the right hand side would
otherwise have a $\delta (x)$
singularity not present on the left.  One also sees that the behavior
of $G \left( x^2 , \varepsilon^2 \right)$ is sufficiently soft as $x
\rightarrow 0$, so that the derivatives in (2.8) can be freely
integrated by parts without generating a short distance surface term.

We now introduce the dimensional constant $M^2$ and separate (2.8)
into the two terms
$$
{1 \over {(x^2 + \varepsilon^2)^2}} = - {1 \over 4} \sq {{\ln \left(
x^2 + \varepsilon^2 \right) M^2} \over x^2} + {1 \over 4} \ln
\varepsilon^2 M^2 \sq {1 \over x^2} \ . \eqnn{2.9}
$$
Here $\sq$ can be interpreted as operating
to the right or left.  Note that
$$
\sq {1 \over x^2} = - 4 \pi^2 \delta (x) \ . \eqnn{2.10}
$$
As $\varepsilon \rightarrow 0$ we obtain
$$
{1 \over {\left( x^2 + \varepsilon^2 \right)^2}} \rightarrow - {1
\over 4} \dsq {{\ln x^2 M^2} \over x^2} - \pi^2 \ln \varepsilon^2 M^2
\delta (x) \ , \eqnn{2.11}
$$
in which we
have introduced the notation $\dsq$ to indicate that the derivative
{\fit must\/} now be interpreted as acting to the left in integrals,
since ill defined singularities would otherwise be obtained.

\vbox{ \ruline {\fbx See figure 2 of Reference [1]}
\item{Fig.~2:}{\fit Diagrams which contribute to $\Gamma^{(4)}
(x_1, x_2, x_3, x_4)$ in $\phi^4$ theory.} \ruline }

To see
the consistency of this interpretation, and its compatibility with
Step 2 of the differential renormalization method let us compute the
Fourier transform of (2.9) and compare with that of (2.11).  The
transform of the left side of (2.9) is easily evaluated using
parametric differentiation,
$$ \eqalign{
\int d^4 x \, e^{ip \cdot x} {1 \over {\left( x^2 + \varepsilon^2
\right)^2}} &= - {1 \over {2 \varepsilon}} {\partial \over {\partial
\varepsilon}} \int d^4 x \, e^{ip \cdot x} {1 \over {(x^2 +
\varepsilon^2)}} \cr
{} &= - {{2 \pi^2} \over \varepsilon} {\partial \over {\partial
\varepsilon}} \left( {\varepsilon \over p} K_1 (\varepsilon p) \right)
\cr
{} &= 2 \pi^2 K_0 (\varepsilon p) \cr
{} &= \pi^2 p^2 \left\{ {2 \over p^2} \left[ K_0 (\varepsilon p) + \ln
\varepsilon M \right] - {{\ln \varepsilon^2 M^2} \over p^2} \right\}
\ . \cr
} \eqnn{2.12}
$$
We have taken the simple exact result in the third line, and rewritten
it with a mass scale $M$ introduced, so that the final form
corresponds to the Fourier transform of the two terms on the right
side of (2.9) with $\sq$ interpreted as the factor $- p^2$.
The limiting form of this result as $\varepsilon \rightarrow 0$,
namely,
$$
- \pi^2 \ln p^2 / \overline M^2 - \pi^2 \ln \varepsilon^2 M^2
\eqnn{2.13}
$$
coincides with the Fourier transform of the right side of (2.11).

The physical interpretation of these manipulations can be seen by
substitution of (2.11) in (2.3) which then reads,
$$ \eqalign{
\Gamma_f^b (x_i , \varepsilon ) &\longrightarrow - 2g^2 \delta_{12}
\delta_{34} \dsq {{\ln x_{13}^2 M^2} \over x_{13}^2} + \perm{2} \cr
{} &\qquad - 24 \pi^2 g^2 \delta_{12} \delta_{13} \delta_{14} \ln
\varepsilon^2 M^2 \cr
{} &= \Gamma_f^r (x_i , M) - 24 \pi^2 g^2 \delta_{12} \delta_{13}
\delta_{14} \ln \varepsilon^2 M^2 \ . \cr } \eqnn{2.14}
$$
This is the sum of the renormalized amplitude of [1] plus a local term
with singular coefficient $\ln \varepsilon^2 M^2$ which is removed by
adding the counterterm
$$
- \pi^2 g^2 \ln \varepsilon^2 M^2 \phi^4 \eqnn{2.15}
$$
to the Lagrangian (2.1).  One can also use (2.13) to obtain the same
interpretation in momentum space, and the renormalized amplitude,
obtained from the first term of (2.13), agrees with the formal
partial integration rule of Step 2.

The other 1-loop diagram is the tadpole contribution $b$ to the
2-point function.  Using the damped propagator (1.2) one finds the
bare contribution to the 1PI 2-point function
$$
\Gamma_b^b (x-y, \varepsilon) = {{2g} \over \varepsilon^2} \delta
(x-y) \ .  \eqnn{2.16}
$$
This can be cancelled immediately by adding the mass counterterm
$$
- {g \over \varepsilon^2} \phi^2 \eqnn{2.17}
$$
to (2.1).  Insertions of the tadpole in other diagrams are also
cancelled by the counterterm graph from (2.17), and higher order
tadpoles can be cancelled similarly.  For this reason we do not
consider tadpole diagrams further in this paper.

The main purpose of this section was to show in 1-loop order using the
cutoff procedure that the bare amplitude can be expressed for small
$\varepsilon$ as the renormalized amplitude of [1] plus a singular
term in $\varepsilon$ which can be absorbed by coupling
renormalization, and that this term corresponds closely to the
singular surface term neglected in the formal partial integration
prescription of differential renormalization.

\sect{3.\ \ 2-Loop Amplitudes}

We now apply the cutoff method of calculation to 2-loop diagrams,
where we shall encounter some new features.

Graph {\ftt c\/} is the first non-trivial contribution to the 2-point
function.
Its bare amplitude is
$$
\Gamma_c^b (x, \varepsilon) = - {{2g^2} \over {3 \pi^2}} \, {1 \over
{(x^2 + \varepsilon^2)^3}} \ .  \eqnn{3.1}
$$
To handle this we simply differentiate the identity (2.8), obtaining
$$
{1 \over {(x^2 + \varepsilon^2)^3}} = - {1 \over {32}} \sq \sq {{\ln
(x^2 + \varepsilon^2) / \varepsilon^2} \over x^2} + {{3 \varepsilon^2}
\over {(x^2 + \varepsilon^2)^4}} \ .  \eqnn{3.2}
$$
The second term in (3.2) is a function whose limit as $\varepsilon
\rightarrow 0$ vanishes for $x \not = 0$, but is singular for
$\varepsilon = 0$.  One suspects that its limiting form is a
distribution, and this may be confirmed by studying the integral
$$
\int {{d^4 x \, f(x) 3 \varepsilon^2} \over {(x^2 + \varepsilon^2)^4}}
= {3 \over \varepsilon^2} \int {{d^4 y \, f(\varepsilon y)} \over
{(y^2 + 1)^4}} \ , \eqnn{3.3}
$$
where $f(x)$ is a smooth function which is damped at large distances.
After the change of variables $x = \varepsilon y$, one can evaluate
the integral for small $\varepsilon$ by expanding $f(\varepsilon y)$
in a Taylor series through second order, and noticing that the
contribution of the remainder is a convergent integral at large
distance which vanishes as $\varepsilon \rightarrow 0$.  After
explicit evaluation of two elementary integrals one finds the result
$$
\int {{d^4 x \, f(x) 3 \varepsilon^2} \over {(x^2 + \varepsilon^2
)^4}} \rightarrow {\pi^2 \over {2 \varepsilon^2}} f(0) + {\pi^2 \over
8} \sq f(0) + {\cal O} (\varepsilon ) \ , \eqnn{3.4}
$$
which is equivalent to the statement
$$
\lim_{\varepsilon \to 0} {{3 \varepsilon^2} \over {(x^2 +
\varepsilon^2)^4}} \rightarrow {\pi^2
\over {2 \varepsilon^2}} \delta (x) + {\pi^2 \over 8} \sq \delta (x)
\ . \eqnn{3.5}
$$
Such ``representations of distributions'' appear frequently in our
work, and they are collected systematically in the Appendix.

The final step in the treatment of (3.2) is to introduce the scale $M$
in the first term, obtaining
$$
\sq \sq {{\ln (x^2 + \varepsilon^2) / \varepsilon^2} \over x^2}
\rightarrow \ \dsq \dsq {{\ln x^2 M^2} \over x^2} + 4 \pi^2 \ln
\varepsilon^2 M^2 \sq \delta (x) \eqnn{3.6}
$$
by steps similar to those leading from (2.8) to (2.11).  We now
combine (3.4) and (3.6) and insert the result in (3.1) obtaining
$$
\Gamma_c^b (x, \varepsilon) \rightarrow {g^2 \over {48 \pi^2}} \dsq
\dsq {{\ln x^2 M^2} \over x^2} + {g^2 \over {12}} \left( \ln
\varepsilon^2 M^2 - 1 \right) \sq \delta (x) - {g^2 \over {3
\varepsilon^2}} \delta (x) \ . \eqnn{3.7}
$$
The first term is the renormalized amplitude $\Gamma_c^r (x,M)$ of [1]
while the last two terms can be cancelled by wave function and mass
counterterms.

We now turn to the 4-point function, for which graphs {\ftt g\/} and
{\ftt h\/}
contribute in 2-loop order.  The bare amplitude for graph {\ftt g\/} is
$$
\Gamma_g^b ( x_i , \varepsilon ) = - {{4 g^3} \over \pi^2} \delta_{12}
\delta_{34} I_\varepsilon (x_{13}) + \perm{2} \ , \eqnn{3.8}
$$
where $I_\varepsilon (x)$ is the convolution integral evaluated
exactly in (A.2).  The limiting form for small $\varepsilon$ is
given by (A.4), and we rewrite it as
$$
I_\varepsilon (x) \rightarrow - {\pi^2 \over 4} \dsq {{\ln^2 x^2 M^2}
\over x^2} + {\pi^2 \over 2} \ln \varepsilon^2 M^2 \left[ \dsq {{\ln
x^2 M^2} \over x^2} + 4 \pi^2 \ln \varepsilon^2 M^2 \delta (x) \right]
- \pi^4 \ln^2 \varepsilon^2 M^2 \delta (x) \ .\eqnn{3.9}
$$
When inserted in (3.8) the first term gives the renormalized amplitude
$\Gamma_g^r$, and the second term is proportional to the limiting form of
$\Gamma_f^b (x_i , \varepsilon)$ in (2.12) times the singular
coefficient $\ln \varepsilon^2 M^2$.  This non-local divergence will
be cancelled, as we show systematically in Section 5, by the bubble
graph generated by the coupling counterterm (2.15).  The last term in
(3.9) can be cancelled by an order $g^3 \ln^2 \varepsilon^2 M^2
\phi^4$ counterterm in the Lagrangian.  The amplitude obtained after
insertion of (3.9) in (3.8) is presented in Table 2.

Graph {\ftt h\/} requires a longer discussion because it is the first
diagram whose basic form is triangular.  The bare amplitude is
$$
\Gamma_h^b (x_i , \varepsilon) = - {{8 g^3} \over \pi^2} \delta_{12} {1
\over {x_{13}^2 + \varepsilon^2}} \, {1 \over {x_{14}^2 +
\varepsilon^2}} \, {1 \over {(x_{34}^2 + \varepsilon^2)^2}} + \perm{5}
\ . \eqnn{3.10}
$$

We let $x_{14} = x$, $x_{34} = y$.  Using (2.9) or (A.10), and (2.10)
together with the antisymmetric derivative identity
$$
A \sq B = \partial_\mu \left( A {\lrhup{\partial}}_\mu
B \right) + B \sq A \ , \eqnn{3.11}
$$
one sees that

$$ \eqalign{
{1 \over {(x-y)^2 + \varepsilon^2}} {1 \over {(x^2 + \varepsilon^2)}}
&{1 \over {(y^2 + \varepsilon^2)^2}} = - {1 \over 4} {\partial \over
{\partial y_\mu}} \left[ {1 \over {(x-y)^2 + \varepsilon^2 }} {1 \over
{x^2 + \varepsilon^2}}
{{\lrhup{\partial}} \over {\partial y_\mu}} {{\ln (y^2 +
\varepsilon^2) M^2} \over y^2} \right] \cr
{} &- {1 \over 4} \sq {1 \over {(x-y)^2 + \varepsilon^2}} {1 \over
{(x^2 + \varepsilon^2)}} {{\ln (y^2 + \varepsilon^2) M^2} \over y^2}
\cr
{} &+ \pi^2 \delta (y) \ln M^2 \varepsilon^2 \left[ {1 \over 4} \sq
{{\ln M^2 ( x^2 + \varepsilon^2 )} \over x^2} + \pi^2 \ln M^2
\varepsilon^2 \delta (x) \right] \ . \cr
} \eqnn{3.12}
$$
Note that (2.9) was used both to replace $1 / (y^2 + \varepsilon^2)^2$
on the left side of (3.12) and to replace $1 / (x^2 +
\varepsilon^2)^2$ in the last term.  The first term is regular as
$\varepsilon \rightarrow 0$ provided we understand that ${\dpar \over
{\partial y_\mu}}$ must be integrated by parts.  A similar remark
applies to $\ln M^2 (x^2 + \varepsilon^2)$ term in the last line.

The second term in (3.12) is an example of something we call a
triangular structure.  To study its limit as $\varepsilon \to 0$, we
use the simple identities
$$
{1 \over {(x-y)^2 + \varepsilon^2}}
= {1 \over {(x-y)^2}} - {\varepsilon^2 \over {(x-y)^2
\left[ (x-y)^2 + \varepsilon^2 \right]}} \eqnn{3.13}
$$

$$
\sq {1 \over {(x-y)^2 + \varepsilon^2}} = - 4 \pi^2 \delta (x-y) - \sq
{\varepsilon^2 \over {(x-y)^2 \left[ (x-y)^2 + \varepsilon^2 \right]}} \
. \eqnn{3.14}
$$
Using (3.14) the second term in (3.12) can be written as
$$
\pi^2 \delta (x-y) {{\ln (x^2 + \varepsilon^2) M^2} \over {x^2 \left(
x^2 + \varepsilon^2 \right) }} + {1 \over 4} \sq {\varepsilon^2 \over
{(x-y)^2 \big( (x-y)^2 + \varepsilon^2 \big) }}  {{\ln \left( y^2 +
\varepsilon^2 \right) M^2} \over {\left( x^2 + \varepsilon^2 \right)
y^2 }} \ . \eqnn{3.15}
$$
The limiting form of the first term is obtained from (A.11a) which is
essentially a differential identity.

We claim that the limiting form of the second term is that of an eight
dimensional delta function $- {1 \over 4} C \delta (x) \delta (y)$
corresponding, after insertion in (3.10), to another two-loop order
coupling
constant counterterm.  This claim can be verified by studying the
integral of the term in question with a test function $f(x,y)$.  After
scaling variables, $x \to \varepsilon x$, $y \to \varepsilon y$, one
sees that the limiting contribution involves only $f(0,0)$, and that
the constant $C$ is given by
$$
C = \int {{d^4 x \, d^4 y} \over {\left( x^2 + 1 \right) y^2 }} \sq
\left( {1 \over {(x-y)^2 \big( (x-y)^2 + 1 \big)}} \right)
\left( \ln \varepsilon^2 M^2 + \ln (y^2 + 1) \right) \eqnn{3.16}
$$ $$
= - 4 \pi^4 \left( \ln \varepsilon^2 M^2 - B \right)
$$
where the first integral becomes trivial after partial integration of
$\sq_y$.  The evaluation of the second integral is discussed briefly
in the Appendix, see (A.13), and the result of the argument beginning
with (3.13) is given in (A.12b). The identities (3.13) and (3.14) are
useful because the product terms fall off fast enough in the infrared so
that their contribution can be obtained by scaling arguments.

It is well worth noting that when scaling arguments are used in the
study of integrals involving test functions, the question of the limit
as $\varepsilon \to 0$ is effectively transferred to the question of
the behavior of the large $x,y$ behavior of integrals over the scaled
variables.  It is a correct rule of thumb, which can be verified by
more careful limiting arguments, that the limiting contribution of a
term in the bare amplitude is a $\delta$-function or product of
$\delta$-functions, if the integral determining the naive coefficient of
the $\delta$-functions is infrared convergent as is the case for $B$
and $C$ in (3.16).

The results (A.11a) and (A.12b) are now combined with the simple
limits of the first and third terms of (3.12) to obtain the limiting
form of $\Gamma_h^b$ given in Table 2.  Again one finds the
renormalized amplitude of [1] plus singular terms to be cancelled by
counterterms.


\sect{4.\ \ 3-Loop Diagrams}

We now continue the program of the last two sections and study the
limit as $\varepsilon \rightarrow 0$ of cutoff amplitudes for the
3-loop graphs shown in Figs.~1 and 2.  It is worth emphasizing that
bare amplitudes are independent of the mass scale $M$,
$$
M {\partial \over {\partial M}} \Gamma^b (x_i, \varepsilon) = 0 \ ,
\eqnn{4.1}
$$
because $M$ is introduced only to separate $\Gamma^b (x_i ,
\varepsilon)$ into regular and singular terms.  Thus one can use the
property (4.1) as a check on the intermediate steps of the calculation
of a complicated amplitude.  The same mass scale is used in all
diagrams in order to agree with the renormalization scheme of [1].

Graph {\ftt d\/} is the only contribution to the 2-point function at
3-loop order.  Its bare amplitude is
$$
\Gamma_d^b (x, \varepsilon) = {g^3 \over \pi^4} {1 \over {x^2 +
\varepsilon^2}} \int d^4 u {1 \over {(u^2 + \varepsilon^2)^2}} {1
\over {\big( (u-x)^2 + \varepsilon^2 \big)^2}} \ .  \eqnn{4.2}
$$
We now use the result (A.2) for the convolution, except that we
compute $\sq$ acting on (A.2). 
We obtain,
$$
\Gamma_d^b (x, \varepsilon) = - {{2g^3} \over \pi^2} {1 \over {x^2 +
\varepsilon^2}} \left\{ {{(x^2 + 2 \varepsilon^2 ) \ln \left( x^2 + 2
\varepsilon^2 - |x| \sqrt {x^2 + 4 \varepsilon^2} \right) /2
\varepsilon^2} \over {|x|^3 {\sqrt {x^2 + 4 \varepsilon^2}}^3 }} + {1
\over {x^2 (x^2 + 4 \varepsilon^2 )}} \right\} \ . \eqnn{4.3}
$$
Inspired by the form of the renormalized amplitude, we compare (4.3)
and
$$
- {g^3 \over {32 \pi^2}} \sq \sq {{\ln^2 (x^2 +
\varepsilon^2)/\varepsilon^2 + 3 \ln (x^2 + \varepsilon^2) /
\varepsilon^2} \over x^2} =
$$ $$
- {{2g^3} \over {\pi^2 (x^2 +
\varepsilon^2)}} \left\{ {{(x^2 - 2 \varepsilon^2) \ln (x^2 +
\varepsilon^2) / \varepsilon^2} \over {(x^2 + \varepsilon^2)^3}} + {x^2
\over {(x^2 + \varepsilon^2)^3}} \right\} \ . \eqnn{4.4}
$$
The difference between (4.3) and (4.4) again involves a representation
of the distributions $\delta(x)$ and $\sq \delta (x)$, as one can
verify by integration with a smooth $f(x)$ as in (3.3--3.4).  One can
then write

$$
\Gamma_d^b (x, \varepsilon) = -{g^3 \over {32 \pi^2}} \sq \sq {{\ln^2
(x^2 + \varepsilon^2) / \varepsilon^2 + 3 \ln (x^2 + \varepsilon^2) /
\varepsilon^2} \over x^2} - {{2g^3} \over {\pi^2 \varepsilon^2}} D_1
\delta (x) - {g^3 \over {4 \pi^2}} D_2 \sq \delta (x) + {\cal O} (
\varepsilon ) \ , \eqnn{4.5}
$$
where $D_1$ and $D_2$ are the purely numerical values of the following
integrals,

$$
D_1 = \int d^4 x {1 \over {x^2 + 1}} \left\{
{{(x^2 + 2) \ln \left( x^2 + 2 - |x| \sqrt
{x^2 + 4} \right) / 2} \over {|x|^3 {\sqrt
{x^2 + 4}}^3 }} + {1 \over {x^2 ( x^2 + 4
)}} \right.
$$ $$
\left. - {{(x^2 - 2) \ln (x^2 + 1 )}
\over {(x^2 + 1 )^3 }} - {x^2 \over {(x^2 +
1)^3}} \right\} \ ,
$$
$$
D_2 = \int d^4 x {x^2 \over {x^2 + 1}} \left\{ {{(x^2 + 2)
\ln \left( x^2 + 2 - |x| \sqrt {x^4 + 4
} \right) / 2} \over {|x|^3 {\sqrt {x^2 + 4
}}^3 }} + {1 \over {x^2 (x^2 + 4)}} \right.
$$ $$
\left. - {{(x^2 - 2 ) \ln (x^2 + 1 )
} \over {(x^2 + 1 )^3}} - {x^2 \over {(x^2 +
1 )^3}} \right\} \ .  \eqnn{4.6}
$$
The logarithmic terms in (4.5) may now be separated into regular and
singular terms in $\varepsilon$ after introduction of the mass scale
$M$.  One then finds the limiting form given in Table 1.  In this
expression $D_1$ and $D_2$ are the coefficients of finite
counterterms at 3-loop order.  These terms become relevant to the
cancellation of divergences only at the 4-loop level, so the integrals
(4.6) need not be evaluated.

\vskip 2pc

\ni {\fbx Table 1.}\ \ {\fit Bare cutoff amplitudes for graphs
contributing
to the 1PI 2-point function.  The subscripts denote the graphs shown
in Fig.~1, and the tadpole graph {\ftt b\/} is omitted for reasons
discussed in Sec.~2.  The first term in each entry is the renormalized
amplitude obtained in [1], and this is followed by cutoff dependent
terms in the limit of small $\varepsilon$.  The numerical constants
$D_1$ and $D_2$ are the values of the integrals in (4.6).}

\vskip 2pc \hrule

$$
\Gamma_{a}(x)= - \sq \delta^4(x)
$$

\ruline

$$
\Gamma_{c}^b (x, \varepsilon) \rightarrow {g^2 \over 48 \pi^2} \dsq
\dsq {{\ln x^2 M^2} \over x^2} - {g^2 \over 12} \left(
1 - \ln \varepsilon^2 M^2 \right) \sq \delta^4 (x)- {g^2 \over 3}
{1 \over \varepsilon^2} \delta^4 (x)
$$

\ruline

$$
\Gamma_{d}^b (x, \varepsilon) \rightarrow -{g^3 \over {32 \pi^2}}
\dsq \dsq {{\ln^2 x^2 M^2 + 3 \ln x^2 M^2} \over x^2}
$$ $$
+3g \ln \varepsilon^2 M^2 \Gamma_{c} (x, \varepsilon ) - {g^3 \over 8}
\left( \ln^2 \varepsilon^2 M^2 + \ln \varepsilon^2 M^2
+ {2D_2 \over \pi^2} \right) \sq \delta^4 (x)
$$ $$
+ g^3 {{\ln \varepsilon^2 M^2 - 2D_1/\pi^2} \over
{\varepsilon^2}} \delta^4 (x)
$$

\ruline


We now begin our treatment of the 8 graphs which contribute to the
4-point function.  We shall be rather brief in our discussion of the
easier graphs and concentrate on the more difficult ones, namely,
{\ftt j\/}, {\ftt l\/}, {\ftt n\/}, and {\ftt o\/}.

Graph {\ftt i\/} is particularly easy since the bare amplitude
$$
\Gamma_i^b (x_i, \varepsilon) = {{2g^4} \over \pi^4} \delta_{12}
\delta_{34} \int d^4 u \, d^4 v {1 \over {\left[ (x_1 - u)^2 +
\varepsilon^2 \right]^2 \left[ (u-v)^2 + \varepsilon^2 \right]^2
\left[ (v-x_3)^2 + \varepsilon^2 \right]^2}} + \perm{2} \ , \eqnn{4.7}
$$
is a double convolution of factors for which the identity (2.8) may be
used.  As discussed in the Appendix, the limiting form of the
convolution is correctly given by the convolution of the limiting form
(2.11) of each factor.  This leads to the result given in Table 2.

Graph {\ftt j\/} is somewhat more involved.  We start by writing its
bare amplitude,
$$
\Gamma_j^b (x_i, \varepsilon) = + {{4 g^4} \over \pi^4} \delta_{12} {1
\over {x_{13}^2 + \varepsilon^2}} {1 \over {x_{14}^2 + \varepsilon^2}}
\int d^4 u {1 \over {\left[ (x_3 - u)^2 + \varepsilon^2 \right]^2}} {1
\over {\left[ (x_4 - u)^2 + \varepsilon^2 \right]^2}} + \perm{5} \ .
\eqnn{4.8}
$$
We substitute the result of the integral (A.2) and then use (3.11),
which splits the amplitude into two parts,
$$
\Gamma_j^b (x_i , \varepsilon) = - {g^4 \over \pi^2} \delta_{12}
\partial_{3 \mu} \left( {1 \over {x_{13}^2 + \varepsilon^2}} {1 \over
{x_{14}^2 + \varepsilon^2}} {\lrhup{\partial}}_{3 \mu} {{\ln^2
\left( x_{34}^2 + 2 \varepsilon^2 - | x_{34}| \sqrt {x_{34}^2 +
4 \varepsilon^2} \right) / 2 \varepsilon^2} \over x_{34}^2} \right)
$$ $$
- {g^4 \over \pi^2} \delta_{12} {1 \over {x_{14}^2 + \varepsilon^2}}
\sq_3 \left( {1 \over {x_{13}^2 + \varepsilon^2}} \right) {{\ln^2
\left( x_{34}^2 + 2 \varepsilon^2 - |x_{34}| \sqrt {x_{34}^2 + 4
\varepsilon^2} \right) / 2 \varepsilon^2} \over x_{34}^2} + \perm{5} \
. \eqnn{4.9}
$$
We take the limit $\varepsilon \rightarrow 0$ in the first one, and
introduce $M$.  This yields
$$
- {g^4 \over \pi^2} \delta_{12} \partial_{3 \mu} \left( {1 \over
{x_{13}^2 + \varepsilon^2}} {1 \over {x_{14}^2 + \varepsilon^2}} {\lrhup
{\partial}}_{3 \mu} {{\left[ \ln^2 \left( x_{34}^2 + 2 \varepsilon^2 - |
x_{34} | \sqrt {x_{34}^2 + 4 \varepsilon^2} \right) / 2 \varepsilon^2
\right]} \over x_{34}^2} \right) \rightarrow
$$ $$
 - {g^4 \over \pi^2} \delta_{12} \dpar_{3 \mu} \left( {1
\over x_{13}^2} {1 \over
x_{14}^2} {\lrhup{\partial}}_{3} {{\ln^2 x_{34}^2 M^2} \over
x_{34}^2} \right) + {{2g^4} \over \pi^2} \ln \varepsilon^2 M^2
\delta_{12} \dpar_3 \left( {1 \over x_{13}^2} {1 \over x_{14}^2}
{\lrhup{\partial}}_{3 \mu} {{\ln x_{34}^2 M^2} \over x_{34}^2} \right)
$$ $$
- {g^4 \over \pi^2} \delta_{12} \partial_{3 \mu} \left( {1 \over
{x_{13}^2 +
\varepsilon^2}} {1 \over {x_{14}^2 + \varepsilon^2}}
{\lrhup{\partial}}_{3 \mu} {1 \over x_{34}^2} \right) \ln^2 \varepsilon^2
M^2 \ . \eqnn{4.10}
$$
The first term of the R.H.S.~is a piece of the renormalized amplitude
of Ref.~[1].  The second is a piece of $g \ln \varepsilon^2 M^2
\Gamma_h^b (x_i, \varepsilon)$.

The third term requires a little more work. We undo relation (3.11)
recognizing a delta term plus the triangular structure (A.12a),
$$
- {g^4 \over \pi^2} \delta_{12} \partial_{3 \mu} \left( {1
\over {x_{13}^2 + \varepsilon^2}} {1 \over {x_{14}^2 + \varepsilon^2}}
{\lrhup{\partial}}_{3 \mu} {1 \over x_{34}^2} \right) \ln^2 \varepsilon^2
M^2 =
$$ $$
4 g^4 \delta_{12} \delta_{34} {1 \over {\left[ x_{13}^2 +
\varepsilon^2 \right]^2}} \ln^2 \varepsilon^2 M^2 - 4g^4 {{\ln^2
\varepsilon^2 M^2} \over {x_{34}^2 \left[ x_{34}^2 + \varepsilon^2
\right]}} \delta_{12} \delta_{13} + 4 g^4 \pi^2 \delta_{12} \delta_{13}
\delta_{14} \ln^2 \varepsilon^2 M^2 \ . \eqnn{4.11}
$$

Let us now turn our attention to the second term of (4.9), which is a
triangular structure independent of scale $M$.  It will have a
representation of the following form in the limit as $\varepsilon
\rightarrow 0$,
$$
- {g^4 \over \pi^2} \delta_{12} \sq_3 \left( {1 \over {x_{13}^2 +
\varepsilon^2}} \right) {1 \over {x_{14}^2 + \varepsilon^2}} {{\ln^2
\left(
x_{34}^2 + 2 \varepsilon^2 - | x_{34} | \sqrt {x_{34}^2 + 4
\varepsilon^2} \right) / 2 \varepsilon^2} \over x_{34}^2} =
$$ $$
 g^4 \delta_{12} \left\{ \delta_{13} {{f(x_{34}^2)}
\over {x_{34}^2 \left(
x_{34}^2 + \varepsilon^2 \right)}} + \pi^2 B_j \delta_{13}
\delta_{14} \right\} \ , \eqnn{4.12}
$$
where $B_j$ is a numerical constant.  We shall determine the function
$f(x_{34}^2)$ by the trick of comparing the left side of (4.12) with a
similar expression in which the logarithm with complicated argument is
replaced by $\ln^2 (x^2 + \varepsilon^2) / \varepsilon^2$.

It can be checked that the difference between the two expressions
amounts, in the small $\varepsilon$ limit, to a delta function whose
constant coefficient $C_j$ is given by
$$
\pi^2 C_j = \int d^4 x {{\ln^2 \left( x^2 + 2 \varepsilon^2 - |x|
\sqrt {x^2 + 4 \varepsilon^2} \right) / 2 \varepsilon^2 - \ln^2 (x^2 +
\varepsilon^2) / \varepsilon^2} \over {x^2 (x^2 + \varepsilon^2)}} \ .
\eqnn{4.13}
$$
Thus, we have
$$
- {g^4 \over \pi^2} \delta_{12} {1 \over {x_{14}^2 + \varepsilon^2}}
\sq_3 \left( {1 \over {x_{13}^2 + \varepsilon^2}} \right) {{\ln^2
\left( x_{34}^2 + 2 \varepsilon^2 - | x_{34} | \sqrt {x_{34}^2 + 4
\varepsilon^2} \right) / 2 \varepsilon^2} \over x_{34}^2} =
$$ $$
 4 g^4 \delta_{12} \delta_{13} {{\ln^2 (x_{34}^2 + \varepsilon^2) /
\varepsilon^2} \over {x_{34}^2 (x_{34}^2 + \varepsilon^2)}} + g^4
\pi^2 (B_j + C_j) \delta_{12} \delta_{13} \delta_{14} \ . \eqnn{4.14}
$$
We introduce $M$ inside the logarithm,
$$
4 g^4 \delta_{12} \delta_{13} {{\ln^2 (x_{34}^2 + \varepsilon^2 ) /
\varepsilon^2} \over {x_{34}^2 (x_{34}^2 + \varepsilon^2)}} =
$$ $$
 4 g^4
\delta_{12} \delta_{13} {{\ln^2 (x_{34}^2 + \varepsilon^2 ) M^2 - 2
\ln (x_{34}^2 + \varepsilon^2) M^2 \ln \varepsilon^2 M^2 + \ln^2
\varepsilon^2 M^2} \over {x_{34}^2 (x_{34}^2 + \varepsilon^2)}} \ .
\eqnn{4.15}
$$
Notice that the second term of the R.H.S.~is another piece of $g \ln
\varepsilon^2 M^2 \Gamma_h^b (x_i, \varepsilon)$.  The third term
cancels with the second term of (4.11).  Finally, we use (A.11b) to
regularize the first term.  These results are collected in the
amplitude of Table 2, where $b_j = C_j + B_j$.

We now turn to graph {\ftt k\/}, whose bare amplitude reads
$$
\Gamma_k^b (x_i , \varepsilon) =
$$ $$
 {{4 g^4} \over \pi^4} \delta_{12}
\int d^4 u {1 \over {\left[ (x_1 - u)^2 + \varepsilon^2 \right]^2
\left[ (x_3 - u)^2 + \varepsilon^2 \right] \left[ (x_4 - u)^2 +
\varepsilon^2 \right] }} \cdot {1 \over {\left[ x_{34}^2 +
\varepsilon^2 \right]}} + \perm{5} \ . \eqnn{4.16}
$$
This is a convolution of bare amplitudes for the bubble and ice-cream
cone subgraphs.  However, the renormalized amplitude [1] for this
graph is also a convolution of the corresponding renormalized
amplitudes interpreted using formal partial integration.  Since our
earlier results show that the limiting forms of the bare amplitudes
for graphs {\ftt f\/} and {\ftt h\/} are equal to the renormalized
amplitudes with $\dsq$ or $\dpar$ derivatives plus singular terms,
these expressions from Table 2 can simply be inserted in (4.16). One
finds the renormalized convolution amplitude plus terms containing
$\delta^4 (u - x_1)$ which render the $d^4 u$ integrals trivial, and
this leads immediately to the result in Table 2.

We now study graph $\ell$ starting from the bare amplitude
$$
\Gamma_\ell^b (x_i , \varepsilon ) = {8 \over {3 \pi^4}} g^4
\delta_{12} \delta_{34} L (x_{13}) + \perm{2} \eqnn{4.17}
$$
where
$$
L(x) = {1 \over {x^2 + \varepsilon^2}} I(x)
$$ $$
I(x) = \int d^4 u \, d^4 v \, {1 \over {(x-u)^2 + \varepsilon^2}} {1
\over {v^2 + \varepsilon^2}} \left[ {1 \over {(u-v)^2 +
\varepsilon^2}} \right]^3 \ . \eqnn{4.18}
$$
The new problem that arises here is that the integral $I(x)$ is infrared
divergent, as may be seen by fixing $u-v$ and considering the integral
over $u+v$.  The physical reason for this is that the subgraph {\ftt
c\/} contains a mass shift which causes the subsequent integration
over the massless propagators to diverge.  One thus expects an
infrared convergent result only when the mass shift counterterm
insertion from (3.5) or (3.7) is subtracted.  Our treatment will make
this clear.

We insert the identity (3.2) in (4.18), and study separately the two
integrals
$$
I_1 (x) = - {1 \over {32}} \int d^4 u \, d^4 v {1 \over {(x-u)^2 +
\varepsilon^2}} {1 \over {v^2 + \varepsilon^2}} \sq \sq {{\ln \left[
(u-v)^2 + \varepsilon^2 \right] / \varepsilon^2} \over {(u-v)^2}}
\eqnn{4.19}
$$ $$
I_2 (x) = 3 \int d^4 u \, d^4 v \, {1 \over {(x-u)^2 + \varepsilon^2}}
{1 \over {v^2 + \varepsilon^2}} {\varepsilon^2 \over {\left[ (u-v)^2 +
\varepsilon^2 \right]^4}} \ . \eqnn{4.20}
$$
The first convolution integral is infrared finite, essentially because
the Fourier transform of the ${\sq \sq \log}$ term contains a factor of
$p^2$
which amply compensates for the $1 / p^4$ factors in the transforms of
the other two propagators.  Therefore we treat $I_2 (x)$ first.

It is not difficult to verify the following differential identity
$$
{{3 \varepsilon^2} \over {\left( z^2 + \varepsilon^2 \right)^4}} = {1
\over 8} \sq {\varepsilon^2 \over {z^2 \left( z^2 + \varepsilon^2
\right)^2}} + {\pi^2 \over {2 \varepsilon^2}} \delta (z) \eqnn{4.21}
$$
in which the $\delta (z)$ singularity cancels between the two terms.
We insert (4.21) with argument $z \to u-v$ in (4.20).  The
contribution of the first term, called $I_{21} (x)$, is infrared
finite, again because there is a factor $p^2$ in momentum space from
the $\sq$ in (4.21).  The infrared divergence is now isolated in the
contribution $I_{22} (x)$ of the last $\delta (u-v)$ term in (4.21),
and it is clear that this integral will be cancelled completely
when the mass counterterm for subgraph {\ftt c\/} is inserted. See
(3.5).

After partial integration of $\sq_v$ and use of (A.6),
we find that $I_{21} (x)$ can be written as
$$ \eqalign{
I_{21} (x) &= - \int {{d^4 u \, d^4 v} \over {(x-u)^2 +
\varepsilon^2}} {\varepsilon^4 \over {(v^2 + \varepsilon^2)^3}} {1
\over {(u-v)^2 \left[ (u-v)^2 + \varepsilon^2 \right]^2}} \cr
{} &= - \int {{d^4 u \, d^4 v} \over {(x - \varepsilon u)^2 +
\varepsilon^2}} {1 \over {(v^2 + 1)^3}} {1 \over {(u-v)^2 \left[
(u-v)^2 + 1 \right]^2}} \cr
} \eqnn{4.22}
$$
where we have scaled $u \to \varepsilon u$ and $v \to \varepsilon v$
in the last line.  It is legitimate to take the $\varepsilon \to 0$
limit inside the integral because the residual integral is finite.
This gives the simple result
$$
I_{21} (x) = - {C \over x^2} \eqnn{4.23}
$$
with
$$ \eqalign{
C &= \int {{d^4 u \, d^4 v} \over {(v^2 + 1)^3 (u-v)^2 \left[ (u-v)^2
+ 1 \right]^2}} \cr
{} &= \int {{d^4 v} \over {(v^2 + 1)^3}} \int {{d^4 z} \over {z^2 (z^2
+ 1)^2}} \cr
{} &= {1 \over 2} \pi^4 \ . \cr
} \eqnn{4.24}
$$

We are really interested in the contribution of $I_{21} (x)$ to $L
(x)$ in (4.18) which is given by the product
$$
L_{21} (x) = {1 \over {x^2 + \varepsilon^2}} I_{21} (x) \eqnn{4.25}
$$
and it is not correct to say that the limiting form of this product is
obtained simply by inserting the limiting form of (4.23) of $I_{21}
(x)$.  Instead we note the following general structure of $I_{21}
(x)$, namely
$$
I_{21} (x) = - {1 \over 2} \pi^4 {1 \over x^2} F \left( \varepsilon^2
/ x^2 \right) \eqnn{4.26}
$$
which follows simply from (4.22).  Because of the result (4.23), we
know that
$$
\lim_{x^2 \to \infty} F \left( \varepsilon^2 / x^2 \right) = 1 \ .
\eqnn{4.27}
$$
We can therefore write
$$
L_{21} (x) = - {1 \over 2} \pi^4 {1 \over {x^2 + \varepsilon^2}} {1
\over x^2} - {1 \over 2} \pi^4 {1 \over {x^2 + \varepsilon^2}} {1
\over x^2} \left[ F \left( \varepsilon^2 / x^2 \right) - 1 \right]
\eqnn{4.28}
$$
which is an exact representation.  Using (A.9.a) one can see that the
limiting form of the first term is that of the bare amplitude for the
bubble graph {\ftt f\/} plus a $\delta (x)$ term.  Using again a test
function and scaling argument, one can show that the limiting
contribution of the second term in (4.20) is also of the local form
$C^\pr \delta (x)$ where $C^\pr$ is a numerical constant defined by
the infrared convergent integral
$$
C^\pr = - {1 \over 2} \pi^4 \int {{d^4 x} \over {x^2 (x^2 + 1)}}
\left[ F (1 / x^2 ) - 1 \right] \ . \eqnn{4.29}
$$
A more explicit form can be found using (4.22), but is not necessary.

The integral $I_1 (x)$ remains to be studied.  We use the identity
(3.13) for each propagator factor obtaining a representation with four
terms
$$
I_1 (x) = \int d^4 u \, d^4 v \, \left[ {1 \over {(x-u)^2 v^2 }} -
{\varepsilon^2 \over {(x-u)^2 \left[ (x-u)^2 + \varepsilon^2 \right]
v^2}} - {\varepsilon^2 \over {(x-u)^2 v^2 \left( v^2 + \varepsilon^2
\right)}} \right. +
$$ $$
+ \left. {\varepsilon^4 \over {(x-u)^2 \left[ (x-u)^2 + \varepsilon^2
\right] v^2 \left( v^2 + \varepsilon^2 \right) }} \right] \sq \sq
{{\ln \left[ (u-v)^2 + \varepsilon^2 \right] / \varepsilon^2} \over
{(u-v)^2 }} \ .  \eqnn{4.30}
$$
After partial integration of $\sq_u \sq_v$ the first term trivially
becomes
$$
I_{11} (x) = 16 \pi^4 {{\ln \left( x^2 + \varepsilon^2 \right) /
\varepsilon^2} \over x^2} \ .  \eqnn{4.31}
$$

Each of the last 3 terms in (4.30) has the structure ${1 \over x^2} F
\left( \varepsilon^2 / x^2 \right)$ because of dimensional
considerations.  If we can show that $F \left( \varepsilon^2 / x^2
\right)$ vanishes as $x^2 \to \infty$, then the contribution of these
terms to $L (x)$ of (4.18) can be shown to be purely local by a
scaling argument similar to that used for the second term of (4.28).
We next discuss how to establish that $F \left( \varepsilon^2 / x^2
\right)$ vanishes for each of the last 3 terms.

After partial integration of $\sq_v$, the second term can be written
as
$$ \eqalign{
I_{12} (x) &= 4 \pi^2 \varepsilon^2 \int {{d^4 u} \over {(x-u)^2 \left[
(x-u)^2 + \varepsilon^2 \right] }} \sq {{\ln \left[ \left( u^2 +
\varepsilon^2 \right) / \varepsilon^2 \right] } \over u^2 } \cr
{} &= - 16 \pi^2 \varepsilon^2 \int {{d^4 u} \over {(x-u)^2 \left[
(x-u)^2 + \varepsilon^2 \right] \left( u^2 + \varepsilon^2 \right)^2}}
\cr
} \eqnn{4.32}
$$
where (2.8) has been used.  A crude estimate of the asymptotic
behavior gives
$$
I_{12} (x) \sim {\varepsilon^2 \over x^4} \int {{d^4 u} \over {\left(
u^2 + \varepsilon^2 \right)^2}} \ .
$$
This is not quite correct since the residual $u$-integral is
logarithmically infrared divergent, but it means that the correct
falloff is
$$
I_{12} (x) \sim {\varepsilon^2 \over x^4} \ln x^2 / \varepsilon^2 \ .
\eqnn{4.33}
$$
The function $F \left( \varepsilon^2 / \varepsilon^2 \right)$ thus
falls nearly a full power of $\varepsilon^2 / x^2$ faster than
necessary, so we are content with the heuristic argument
above.  The third integral in (4.30), namely $I_{13} (x)$, can be
shown to be equal to $I_{12} (x)$ after partial integration of $\sq_u$
and change of variables, so $I_{13} (x)$ also satisfies (4.33).

The fourth integral $I_{14} (x)$ is more complicated, but it is not
difficult to show that it falls rapidly as $x \to \infty$.  If we
crudely extract the factor $1 / x^4$ and use (2.8) we find
$$ \eqalign{
I_{14} (x) &\sim {\varepsilon^4 \over x^4} \int d^4 u \, d^4 v \, \sq_v
{1 \over {\left[ (u-v)^2 + \varepsilon^2 \right]^2}} {1 \over {v^2
\left( v^2 + \varepsilon^2 \right)}} \cr
{} &\sim {\varepsilon^4 \over x^4} \int d^4 u \, d^4 v \, \vec
\nabla_u {1 \over {\left[ (u-v)^2 + \varepsilon^2 \right]^2}} \cdot
\vec \nabla_v \left( {1 \over {v^2 \left( v^2 + \varepsilon^2
\right)}} \right) \cr
{} &\sim {\varepsilon^4 \over x^4} \int d^4 u^\pr \vec \nabla {1 \over
{\left( u^{\pr 2} + \varepsilon^2 \right)^2}} \cdot \int d^4 v \vec
\nabla \left( {1 \over {v^2 \left( v^2 + \varepsilon^2 \right)}} \right)
\ .  \cr %
} \eqnn{4.34}
$$
The translation of variables $u^\pr = u-v$ is permitted because the
eight dimensional integral is convergent.  Each of the two
four-dimensional convergent integrals in the last line vanishes by
symmetry, and this means that $I_{14} (x)$ actually falls faster than
$1 / x^4$.  This is more than enough to conclude that its contribution
to $L(x)$ is purely local as $\varepsilon \to 0$.

The total contribution of $I_1 (x)$ to $L (x)$ is therefore
$$
L_1 (x) = 16 \pi^4 {{\ln \left( x^2 + \varepsilon^2 \right) /
\varepsilon^2} \over {x^2 \left( x^2 + \varepsilon^2 \right)}} +
C^\dpr \delta (x) \eqnn{4.35}
$$
where the first term comes from (4.31).  We introduce the scale $M$ in
the logarithm and apply the differential identity (A.11a).  Similarly
(A.9a) and (A.10a) are applied to the first term of (4.28).  The
limiting form of the bare amplitude $\Gamma_\ell^b (x_i ,
\varepsilon)$ given in Table 2 is the contribution of these two terms
plus the infrared divergent mass counterterm integral $I_{22} (x)$
discussed below (4.21) and a local triple $\delta$-term.

The bare amplitude of graph {\ftt m\/} is
$$
\Gamma_m^b (x_i , \varepsilon ) = {{4g^4} \over \pi^4}
{1 \over {(x_{12}^2 + \varepsilon^2
)^2 }} {1 \over {(x_{34}^2 + \varepsilon^2 )^2}} {1 \over {x_{13}^2 +
\varepsilon^2 }} {1 \over {x_{24}^2 + \varepsilon^2}} + \perm{5} \ .
\eqnn{4.36}
$$
We now use (2.9) for the bubble subgraph amplitudes which leads to
$$
\Gamma_m^b (x_i) = {g^4 \over {4 \pi^4}} \sq {{\ln (x_{12}^2 +
\varepsilon^2 ) M^2} \over x_{12}^2} \sq {{\ln (x_{34}^2 +
\varepsilon^2 ) M^2} \over x_{34}^2} {1 \over {x_{13}^2 +
\varepsilon^2}} {1 \over {x_{24}^2 + \varepsilon^2}} + \perm{5}
$$ $$
+ g \ln \varepsilon^2 M^2 \Gamma_h^b (x_i , \varepsilon) - g^2 \ln^2
\varepsilon^2 M^2 \Gamma_f^b (x_i , \varepsilon) \ . \eqnn{4.37}
$$
Using (3.11) we obtain the antisymmetric derivative terms in Table 2,
plus the expression
$$
F = {g^4 \over {4 \pi^4}} \sq \left( {1 \over {x_{13}^2 +
\varepsilon^2}} \right) {{\ln (x_{12}^2 + \varepsilon^2 ) M^2} \over
x_{12}^2} \sq {1 \over {x_{24}^2 + \varepsilon^2 }} {{\ln (x_{34}^2 +
\varepsilon^2 ) M^2} \over x_{34}^2} \ . \eqnn{4.38}
$$

The limiting behavior of $F$ can be obtained by a procedure involving
the use of the identity (3.14) in two factors of (4.38), namely those
involving $\sq$ of the cutoff propagators with arguments $x_{13}$ and
$x_{14}$.  A detailed discussion would be lengthy, and since there are
no essentially new techniques involved, we give only a brief
description.  It is convenient to split the product of logarithms in
(4.38) into terms proportional to $\ln^2 \varepsilon^2 M^2$, $\ln
\varepsilon^2 M^2$, and $M$-independent terms, noting that factors
such as $\left( \ln \left[ (x^2 + \varepsilon^2) / \varepsilon^2 \right]
\right) / x^2$ are non-singular as $x \to 0$.  Each of the three terms
above gives rise to four terms coming from the product of the two
identities based on (3.14), and each term can be studied separately in
a straightforward way.  A minor difficulty occurs in the $\ln^2
\varepsilon^2 M^2$ term, because the product $1 / \left( x_{12}^2
x_{34}^2 \right)$ becomes ultraviolet singular when the arguments are
identified in the $\delta (x_{13}) \delta (x_{24})$ term of the
identities (3.14).  to handle this one uses essentially (3.13), namely
$$
{1 \over x_{12}^2} = {1 \over {x_{12}^2 + \varepsilon^2}} +
{\varepsilon^2 \over {x_{12}^2 \left( x_{12}^2 + \varepsilon^2 \right)
}} \ .
$$
The first term cuts off the ultraviolet singularity of the product $1
/ \left( x_{12}^2 x_{34}^2 \right)$.  The contribution of the second
term to the $\ln^2 \varepsilon^2 M^2$ term of (4.38) is then studied,
without use of (3.14), and can be shown to be of the local form
$\delta_{12} \delta_{13} \delta_{14}$ as $\varepsilon \to 0$.

The result of the analysis above is the following formula for the
$\varepsilon \to 0$ limit of $F$:
$$
F \rightarrow 4 g^4 \delta_{13} \delta_{24} {{\ln^2 (x_{12} +
\varepsilon^2 ) M^2} \over {x_{12}^2 (x_{12}^2 + \varepsilon^2) }} -
4g^4 \pi^2 \delta_{13} \delta_{12} \delta_{14} \left( \ln^2
\varepsilon^2 M^2 - 2 B \ln \varepsilon^2 M^2 + B_m \right) \ .
\eqnn{4.39}
$$
All triple $\delta$-terms were obtained by studying the limiting
behavior of integrals with a test function
$h \left( x_{12} , x_{13} , x_{14} \right)$ of the three independent
variables in (4.38).  The purely numerical coefficient $B_m$ comes
from a combination of integrals from various terms in the analysis
above. One would expect that the singularity associated with the bubble
subgraphs of {\ftt m\/} should involve the amplitude of the ice-cream
cone subgraph {\ftt h\/}.
It is therefore important to verify, as we have
done in detail, that the coefficient $B$ in (4.39) is given by the
same integral, see (3.16) or (A.13), that appeared in the original
analysis of graph {\ftt h.\/}  Only then we will have a consistent
cancellation of divergences as $\varepsilon \to 0$ by counterterms.
A further check on (4.39) can be obtained by verifying that the
limiting form of $\Gamma_m^b$ given in Table 2 satisfies (4.1).  The
net coefficient of $B \ln \varepsilon^2 M^2$ is easily seen to cancel.

Graph {\ftt n\/} is moderately complicated. Its bare amplitude is

$$
\Gamma_n^b (x_i , \varepsilon ) = {{8 g^4} \over \pi^4} \delta_{12} {1
\over {x_{14}^2 + \varepsilon^2}} {1 \over {x_{34}^2 + \varepsilon^2}}
\, \cdot
$$ $$
\cdot \, \int d^4 u {1 \over {\left[ (x_1 - u)^2 + \varepsilon^2
\right] \left[ (
x_4 - u)^2 + \varepsilon^2 \right] \left[ (x_3 - u)^2 + \varepsilon^2
\right]^2 }} + \perm{11} \ .  \eqnn{4.40}
$$
We focus attention on the integral, use (2.8), and follow the spirit
of the corresponding steps in [1] to obtain the two terms
$$
\Gamma_n^b (x_i , \varepsilon) = - {{2g^4} \over \pi^4} \delta_{12} {1
\over {x_{14}^2 + \varepsilon^2}} {1 \over {x_{34}^2 + \varepsilon^2}}
\sq_3 \int d^4 u {1 \over {(u^2 + \varepsilon^2) \left( (x_{14} - u)^2
+ \varepsilon^2 \right) (x_{13} - u)^2}} \, \cdot
$$ $$
\ln \left( {{(x_{13} - u)^2 +
\varepsilon^2} \over {(x_{14} - u)^2 + \varepsilon^2}} \right)
+ {{8g^4} \over \pi^2} \delta_{12} {1 \over {(x_{13}^2 +
\varepsilon^2) (x_{14}^2 + \varepsilon^2)}} {{\ln (x_{34}^2 +
\varepsilon^2) / \varepsilon^2} \over {(x_{34}^2 + \varepsilon^2 )^2}}
+ \perm{11} \eqnn{4.41}
$$
as an exact result.  We now use (3.11) to split the first term above
into two parts.  The term with the anti-symmetric derivative is
already finite as $\varepsilon \rightarrow 0$, so we can use the
techniques of [1] to obtain the corresponding term in the renormalized
amplitude.

The second term resulting from use of (3.11) in (4.41)
contains the difference of the
two triangular structures (A.12c) for $n=1$ and (A.12d).
The $u$-integral cancels and one
finds that the total contribution of the first term in
(4.41) has the limiting form
$$
- {g^4 \over \pi^4} \delta_{12} \dpar_{3 \mu} \left( {1 \over
{x_{14}^2 x_{34}^2}} {\lrhup{\partial}}_{3 \mu} \ln {x_{13}^2 \over
x_{14}^2} K (x_{13} , x_{14}) \right) + \perm{11} + g^2 (4B + 4)
\Gamma_f^b (x_i , \varepsilon) \eqnn{4.42}
$$
where
$$
K (x,y) = \int d^4 u {1 \over {u^2 (x-u)^2 (y - u)^2}} \eqnn{4.43}
$$
is the same function introduced in [1].  $K (x,y)$ has an ultraviolet
finite Fourier transform.

The remaining task is to study the second term in (4.41).  This is
straightforward if we use (A.10c) and introduce the scale $M$ to
obtain the exact result
$$
- {g^4 \over \pi^2} \delta_{12} {1 \over {(x_{13}^2 + \varepsilon^2 ) (
x_{14}^2 + \varepsilon^2)}} \sq_3 {{\ln^2 (x_{34}^2 + \varepsilon^2)
M^2 + 2 \ln (x_{34}^2 + \varepsilon^2) M^2} \over x_{34}^2} +
\perm{11}
$$ $$
+ 2g \ln \varepsilon^2 M^2 \Gamma_h^b (x_i , \varepsilon ) - 2g^2
\left( \ln^2 \varepsilon^2 M^2 + 2 \ln \varepsilon^2 M^2 \right)
\Gamma_f^b (x_i , \varepsilon) \ . \eqnn{4.44}
$$
We use (3.11) again, obtaining an ultraviolet finite antisymmetric
derivative term and a triangular structure, which is a combination of
(A.12.a -- A.12.c).
We then use (A.11a) and (A.11b) and assemble our results to complete
the amplitude given in Table 2.

We shall use a different technique to analyze graph {\ftt o\/} in
order to avoid an intractable interplay of internal integrals and
$\varepsilon \rightarrow 0$ limits.  Namely, we will obtain the
renormalized amplitudes plus non-local divergent terms directly, but
we will use (4.1) to obtain the coefficients of scale dependent local
triple $\delta$ terms.  This is a mathematically correct shortcut because the
bare amplitude is independent of $M$.  These considerations determine
the limiting form of amplitude except for a purely numerical multiple
of $\delta_{12} \delta_{13} \delta_{14}$ which is simply a change in
the renormalization scheme of [1].

The bare amplitude of graph {\ftt o\/} is
$$
\Gamma_o^b (x_i , \varepsilon) = {{4 g^4} \over \pi^4} \delta_{12}
\delta_{34} f_o^b (x_{13})
$$ $$
f_o^b (x) = \int {{d^4 u \, d^4 v} \over {\left( (u-v)^2 +
\varepsilon^2 \right)^2}} {1 \over {u^2 + \varepsilon^2}} {1 \over
{v^2 + \varepsilon^2}} {1 \over {(x-u)^2 + \varepsilon^2}} {1 \over
{(x-v)^2 + \varepsilon^2}} \ . \eqnn{4.45}
$$
We regulate the bubble subgraph factor using (2.9) and integrate
$\sq_u$ by parts obtaining the three integrals
$$
f_o^b (x) = - {1 \over 4} \int d^4 u \, d^4 v \, \left\{ \sq {1 \over
{u^2 + \varepsilon^2}} {1 \over {v^2 + \varepsilon^2}} {1 \over
{(x-u)^2 + \varepsilon^2}} {1 \over {(x-v)^2 + \varepsilon^2}} \right.
$$ $$
+ {1 \over {u^2 + \varepsilon^2}} {1 \over {v^2 + \varepsilon^2}} \sq
{1 \over {(x-u)^2 + \varepsilon^2}} {1 \over {(x-v)^2 +
\varepsilon^2}}
$$ $$
\left. + 2 \partial_\mu {1 \over {u^2 + \varepsilon^2}} {1 \over {v^2 +
\varepsilon^2}} \partial_\mu {1 \over {(x-u)^2 + \varepsilon^2}} {1
\over {(x-v)^2}} \right\} {{\ln \left( (u-v)^2 + \varepsilon^2 \right)
M^2} \over {(u-v)^2}} \eqnn{4.46}
$$
plus a contribution to $\Gamma_o^b (x_i , \varepsilon)$ proportional
to the bare amplitude of graph {\ftt g\/}, namely $g \ln \varepsilon^2
M^2 \Gamma_g^b (x_i , \varepsilon)$.

The first two terms of (4.46) have the same form, containing the
triangular
structure (A.12b). We are entitled to use formula (A.12b) inside
the integral provided that $x \not = 0$. To account for a possible
singularity
in this limit, we include a $\delta (x)$ term, with unknown coefficient
$F_0^1 (\varepsilon M)$. Such coefficient will be fixed at the end of
the computation, requiring the amplitude to satisfy equation
(4.1). So we have
$$
\int d^4 u d^4 v \sq \left[ {1 \over {u^2 + \varepsilon^2}} \right] {1
\over {v^2 + \varepsilon^2}} {1 \over {(x-u)^2 + \varepsilon^2}} {1
\over {(x-v)^2 + \varepsilon^2}} {{\ln \left( (u-v)^2 + \varepsilon^2
\right) M^2} \over {(u-v)^2}} \rightarrow
$$ $$
4 \pi^2 \int d^4 v {1 \over {x^2 + \varepsilon^2}} {1
\over {(x-v)^2
+ \varepsilon^2}} {{\ln (v^2 + \varepsilon^2 ) M^2} \over {v^2 (v^2 +
\varepsilon^2)}} + 4 \pi^4 (\ln \varepsilon^2 M^2 - B) {1 \over {[ x^2
+ \varepsilon^2 ]^2}} + F_0^1 (\varepsilon M) \delta (x) \ . \eqnn{4.47}
$$
We regulate the divergent term in the integral using (A.11a),
integrate by parts and regulate again using (A.11a) and (A.11b).  One
thus obtains the following limiting form of the first two integrals in
(4.46)
$$
- {\pi^4 \over {12}} \dsq {{\ln^3 x^2 M^2 + 6 \ln^2 x^2 M^2
+ 12 \ln x^2 M^2} \over x^2 } +
$$ $$
+ \pi^4 \left( - \ln^2
\varepsilon^2 M^2 - 2 \ln \varepsilon^2 M^2 + 2B + 2 \right) \cdot {1
\over {(x^2 + \varepsilon^2 )^2}} + F_0^2(\varepsilon M) \delta (x) \ .
\eqnn{4.48} $$

We now study the last integral in (4.46).  If $x \not = 0$, the
integral actually converges if the limit $\varepsilon \rightarrow 0$
is taken in the integrand, and the result
$$
4 \pi^4 \left[ {{\ln x^2 M^2} \over x^4} + 2 {1 \over
x^4} \right] \eqnn{4.49}
$$
was obtained for this limit in [1] by mathematically correct steps not
requiring formal partial integration.  The role of the cutoff
$\varepsilon$ is therefore to determine the $\delta (x)$ term in
the result of the integral.  We are therefore entitled to assume that
the integral takes the form
$$
4 \pi^4 \left[ {{\ln (x^2 + \varepsilon^2) M^2} \over {x^2
(x^2 + \varepsilon^2)}} + 2 {1 \over {\left( x^2 +
\varepsilon^2 \right)^2}} \right] + F_0^3 (\varepsilon M) \delta (x)
\eqnn{4.50}
$$
as $\varepsilon \rightarrow 0$.  Any changes in the way the
$x^{-4}$ singularities are cut off results only in a change in
the function $F_0^3 (\varepsilon M)$.  See, for
example, (A.9a--A.9b).
The first two terms are then regulated in the standard fashion, using
(A.10a -- A.11a).  We now insert the results (4.48) and (4.50) in (4.46)
and use (4.1) to determine the scale dependent part of $F_0^2
(\varepsilon M) + F_0^3 (\varepsilon M)$.
In this way we obtain the complete limiting form of
the cutoff amplitude, except for a $\delta_{12} \delta_{13}
\delta_{14}$ term whose unknown numerical constant is called $b_0$.

The last graph needed, namely $p$, is primitively divergent.  A special
device
was used to regulate it in Ref.~[1] (see below) and it is particularly
useful
to see that the same renormalized amplitude can be obtained from the
cutoff
procedure.  In this discussion below we will use some arguments from
the
treatment of Ref.~[1] which do not involve the assumption of formal
partial
integration.

The bare amplitude for graph $p$ is
$$
\Gamma^b_p \left( x_i,\varepsilon\right) = {16 g^4\over \pi^4} f
\left( x_{12},
x_{13}, x_{14}, \varepsilon\right)
$$ $$
f\left( x,y,z,\varepsilon\right) = {1\over x^2 + \varepsilon^2} \
{1\over y^2 +
\varepsilon^2} \  {1\over z^2 + \varepsilon^2}\ {1\over \left(
x-y\right)^2+\varepsilon^2}\ {1\over \left( y-z\right)^2 +
\varepsilon^2}\  {1\over
\left( z-x\right)^2 + \varepsilon^2}\ \ .\eqnn{4.51}$$
Because the graph is primitively divergent, it is sufficient
to cut off only
one of the six propagators to  obtain an amplitude with a well-defined
Fourier transform. We therefore add and subtract the product of the
last five
propagators without $\varepsilon$ and write
$$f\left( x,y,z,\varepsilon\right) = {1\over x^2+\varepsilon^2}\
{1\over y^2}\
{1\over z^2}\  {1\over (x-y)^2}\ {1\over \left(y-z\right)^2}\
{1\over\left(
z-x\right)^2} + r\left( x,y,z,\varepsilon\right)\ \ .\eqnn{4.52}$$
It is easy to show by scaling that the limiting form of the
remainder term
$r(x,y,z,\varepsilon)$ is $C\delta(x)\,\delta(y)\,\delta(z)$
where $C$ is given
by an integral which is infrared convergent because the integrand
is a
difference of two terms with the same leading infrared behavior.
We drop this
term henceforth.

We now write
$$f\left( x,y,z,\varepsilon\right) = {1\over \left( x^2 +
\varepsilon^2\right)^2} \
{x^2 + \varepsilon^2\over y^2 z^2 \left( x-y\right)^2
\left( y-z\right)^2 \left(
z-x\right)^2} \eqnn{4.53}$$
where we have imitated the special device of Ref.~[1]
in which the degree of
singularity of the first propagator was artificially
increased.  We drop
the explicit $\varepsilon^2$ term in the numerator of
(4.53) because it is also
shown easily to contribute a finite triple $\delta$-term
as $\varepsilon\to 0$.
We now use (2.8) and study
$$f\left( x,y,z,\varepsilon\right) = - {1\over 4}\sq
\left[{\ln \left(
x^2+\varepsilon^2\right)\big/\varepsilon^2\over x^2}\right]
\cdot {x^2\over y^2 z^2
\left( x-y\right)^2 \left( y-z\right)^2 \left( z-x\right)^2}
\ \
.\eqnn{4.54}$$
The term in brackets is regular as $x\to 0$, so that the box
operator can be
integrated by parts without surface terms in integrals of
(4.54) with smooth
test functions such as the Fourier transform studied in Ref.~[1].
We indicate
this partial integration with $\dsq$, and split the argument of the
log to obtain
$$f\left( x,y,z,\varepsilon\right) = - {1\over 4} \dsq \left[ {\ln
\left( x^2+\varepsilon^2\right)M^2 - \ln \varepsilon^2 M^2\over x^2}
\right]
{x^2\over y^2 z^2 \left( z-y\right)^2 \left( y-z\right)^2 \left( z-x
\right)^2}
\ \ .\eqnn{4.55}$$
If we replace $\ln \left( x^2 + \varepsilon^2\right)M^2
\longrightarrow \ln x^2
M^2$ in the first term, the result is just the renormalized
amplitude of
Ref.~[1], which was shown there to give ultraviolet convergent
integrals with
test functions.  Note that in these integrals the derivatives in
$\dsq$ are applied both to the test function and the second factor in
(4.55).  The replacement made above can be justified by studying
integrals in
which the difference
$${\ln \left( x^2+\varepsilon^2\right) M^2 - \ln x^2 M^2\over x^2}
= {\ln \left(
1 + \varepsilon^2\big/x^2\right)\over x^2}\eqnn{4.56}$$
is inserted in (4.55).  Scaling arguments show that such integrals
vanish as
$\varepsilon\to 0$, because of the structure found in Ref.~[1] for the
$y-z$
subintegrals as $x\to 0$ and because the integrand vanishes faster as
$x\to\infty$ than that of (4.55) itself.

In the second term of (4.55), we reverse the partial integration
and pick up a
surface term which is exactly that obtained in the rigorous
derivation of
(2.10) from Green's identity.  The second term is then
$$- \pi^2 \delta(x) \lim\limits_{x\to 0} {x^2\over y^2 z^2
\left( x-y\right)^2
\left( y-z\right)^2 \left( z-x\right)^2} = - 6\pi^4 \zeta(3)\,
\delta (x)\,
\delta(y)\, \delta(z) \eqnn{4.57}$$
where $\zeta(s)$ is the Riemann $\zeta$-function as found from
a study of the
limit as $x\to 0$ in Ref.~[1].

The result of this analysis is given in Table~2 in which we have
added
$-96 \pi^2 g^4 b_p\,\delta(x)\,\delta(y)\,\delta(z)$ to account
for the
finite triple $\delta$-terms dropped above.

\vskip 2pc


\def\tripledelta{\delta_{12} \delta_{13} \delta_{14} }

\ni {\fbx Table 2.}\ \ {\fit The limiting form of bare amplitudes for
graphs contributing to the 1PI 4-point function.  Subscripts denote the
graphs shown in Fig.~2.  The first term in each entry is the
renormalized amplitude obtained in [1], including the number of
permutations of external points $x_i$ required to obtain the full
contribution of a given graph.  Cutoff dependent terms in the limit of
small $\varepsilon$ are then given.  The constant $B$ is given in
(A.13).  Numerical constants $b_\ell$, $b_m$, $\ldots$, in 3-loop
graphs can be expressed as similar integrals, but their specific form
is not required.}

\vskip 2pc \hrule

$$
\Gamma_{e} (x_i) = - 16 \pi^2 g \delta_{12}
\delta_{13} \delta_{14}
$$

\ruline

$$
\Gamma_{f}^b (x_i , \varepsilon ) \to -2g^2 \delta_{12}
\delta_{34} \dsq {{\ln x_{13}^2 M^2} \over {x_{13}^2}}
+ \perm{2}
$$ $$
-24 \pi^2 g^2 \ln \varepsilon^2 M^2 \tripledelta
$$

\ruline 

$$
\Gamma_{g}^b (x_i , \varepsilon ) \to g^3 \delta_{12} \delta^4
(x_{34}) \dsq {{\ln^2 x_{13}^2 M^2} \over {x_{13}^2}} + \perm{2}
$$ $$
+g \ln \varepsilon^2 M^2  \Gamma_{f}^b (x_i , \varepsilon)
$$ $$
+ 12 \pi^2 g^3 \ln^2 \varepsilon^2 M^2 \tripledelta
$$

\ruline

$$
\Gamma_{h}^b (x_i , \varepsilon) \to {{2 g^3} \over \pi^2} \Biggl[
{\pi^2 \over 2} \delta_{12} \delta_{13} \dsq
{{\ln^2 x_{34}^2 M^2 + 2 \ln x_{34}^2 M^2} \over x_{34}^2}
$$ $$
+\delta_{12} {\dpar \over {\partial x_3^\mu}} \left( {1
\over {x_{14}^2 x_{13}^2}} {\lrhup{\partial} \over {\partial x_3^\mu}}
{{\ln x_{34}^2 M^2} \over x_{34}^2} \right) \Biggr] + \perm{5}
$$ $$
+2g \ln \varepsilon^2 M^2 \Gamma_f^b (x_i , \varepsilon)
$$ $$
+48 \pi^2 g^3 \left( {1 \over 2} \ln^2 \varepsilon^2 M^2
+ \ln \varepsilon^2 M^2 - 1 - B \right) \tripledelta
$$

\ruline

$$
\Gamma_i^b (x_i , \varepsilon) \to - {g^4 \over 2} \delta_{12}
\delta_{34} \left( \dsq {{\ln^3 x_{13}^2 M^2}
\over {x_{13}^2}} -16 \pi^2 \big(1 - \zeta (3) \big)
\delta_{13} \right) + \perm{2} \, .)
$$ $$
+{3g \over 2} \ln \varepsilon^2 M^2 \Gamma_{g}^b (x_i , \varepsilon)
- {3g^2 \over 4} \ln^2 \varepsilon^2 M^2 \Gamma_{f}^b
(x_i , \varepsilon )
$$ $$
- 6 \pi^2 g^4 \ln^3 \varepsilon^2 M^2 \tripledelta
$$

\ruline

$$
\Gamma_{j}^b (x_i , \varepsilon ) \to - {g^4 \over \pi^2} \delta_{12}
\Bigg[ {\dpar \over {\partial x_3^\mu}} \left( {1 \over
x_{13}^2 x_{14}^2} {\lrhup{\partial} \over {\partial x_3^\mu}}
{{\ln^2 x_{34}^2 M^2} \over {x_{34}^2}} \right)
$$ $$
+ {\pi^2 \over 3} \delta_{13} \dsq {{\ln^3 x_{34}^2 M^2 + 3
\ln^2 x_{34}^2 M^2 + 6 \ln x_{34}^2 M^2} \over {x_{34}^2}}
\Bigg] + \perm{5}
$$ $$
+g \ln \varepsilon^2 M^2 \Gamma_{h}^b (x_i , \varepsilon ) - g^2
\ln^2 \varepsilon^2 M^2  \Gamma_{f}^b (x_i , \varepsilon )
$$ $$
- \pi^2 g^4 \left( 8 \ln^3 \varepsilon^2 M^2 + 24 \ln^2
\varepsilon^2 M^2 - 48B \ln \varepsilon^2 M^2 - 12 -6
b_j \right) \tripledelta
$$

\ruline 

$$
\Gamma_{k}^b (x_i , \varepsilon ) \to {g^4 \over {4 \pi^2}} \left\{
\delta_{12} \dsq_{x_1} \int d^4 u {{\ln (u-x_1)^2 M^2} \over
{(u-x_1)^2}} \cdot \right.
$$ $$
\left[ {\dpar \over {\partial {x_3}^\mu}} \left( {1 \over {(u -
x_3)^2}} {1 \over {(u - x_4)^2}} {\lrhup{\partial} \over {\partial
x_3^\mu}}
{{\ln x_{34}^2 M^2} \over {x_{34}^2}} \right) + \right.
$$ $$
\left. \left. + {\pi^2 \over 2} \delta^4 (u - x_3) \dsq_{x_4} {{\ln^2
(u
- x_4)^2 M^2 + 2 \ln (u - x_4)^2 M^2} \over {(u - x_4)^2}} \right]
\right\} + \perm{5}
$$ $$
+ {g \over 2} \ln \varepsilon^2 M^2 \Gamma_{h}^b (x_i , \varepsilon )
+ 2g \ln \varepsilon^2 M^2 \Gamma_{g}^b (x_i , \varepsilon)
$$ $$
- g^2 \left( {3 \over 2} \ln^2 \varepsilon^2 M^2 +\ln
\varepsilon^2 M^2  -1 - B \right) \Gamma_{f}^b (x_i , \varepsilon)
$$ $$
- 24 \pi^2 g^4 \left( {{\ln^3 \varepsilon^2 M^2} \over 2} + \ln^2
\varepsilon^2 M^2 - \ln \varepsilon^2 M^2 - B\ln \varepsilon^2 M^2
\right) \tripledelta
$$

\ruline

$$
\Gamma_\ell^b (x_i , \varepsilon) \to {g^4 \over 6} \delta_{12}
\delta_{34} \dsq {{\ln^2 x_{13}^2 M^2+ 2 \ln x_{13}^2 M^2}
\over {x_{13}^2}} + \perm{2}
$$ $$
+ {g^2 \over 6} \left( \ln \varepsilon^2 M^2 -1\right) \Gamma_{f}^b
(x_i , \varepsilon)
$$ $$
- {{4 g^4} \over {3 \pi^2 \varepsilon^2}} \delta_{12}
\delta_{34} {1 \over x_{13}^2} \int d^4 u {1 \over {(x_1-u)^2
(x_3-u)^2}} + \perm{2}
$$ $$
+2 \pi^2 g^4 \left( \ln^2 \varepsilon^2 M^2 +2\ln \varepsilon^2
M^2 - 2 + b_l \right) \tripledelta
$$

\ruline

$$
\Gamma_{m}^b (x_i , \varepsilon ) \to {g^4 \over 4 \pi^4}
\Biggl[ {\dpar \over {\partial x_2^\mu}} {\dpar \over {\partial
x_3^\nu}} \left( {1 \over x_{24}^2} {\lrhup{\partial} \over
{\partial x_2^\mu}}
{{\ln x_{12}^2 M^2} \over {x_{12}^2}} \right) \left(
{1 \over {x_{13}^2}} {\lrhup{\partial} \over {\partial x_3^\nu}}
{{\ln x_{34}^2 M^2} \over {x_{34}^2}} \right)
$$ $$
- 4 \pi^2 \delta_{24} {{\ln x_{14}^2 M^2} \over {x_{14}^2}}
{\dpar \over {\partial x_3^\mu}} \left( {1 \over {x_{13}^2}}
{\lrhup{\partial} \over {\partial x_3^\mu}} {{\ln x_{34}^2 M^2}
\over {x_{34}^2}} \right)
$$ $$
- 4 \pi^2 \delta_{13} {{\ln x_{34}^2 M^2} \over
{x_{34}^2}} {\dpar \over {\partial x_2^\mu}}\left( {1 \over
{x_{24}^2}} {\lrhup{\partial} \over {\partial x_2^\mu}}
{{\ln x_{12}^2 M^2} \over {x_{12}^2}} \right)
$$ $$
- {{4 \pi^4} \over 3} \delta_{24} \delta_{13}
\dsq {{\ln^3 x_{14}^2 M^2 + 3 \ln^2 x_{14}^2 M^2
+ 6 \ln x_{14}^2 M^2} \over {x_{14}^2}} \Biggr] + \perm{5}
$$ $$
+g \ln \varepsilon^2 M^2 \Gamma_{h}^b (x_i , \varepsilon) -g^2 \ln^2
\varepsilon^2 M^2 \Gamma_{f}^b (x_i , \varepsilon)
$$ $$
- 48 \pi^2 g^4 \left( {1 \over 6} \ln^3 \varepsilon^2 M^2 +
{{\ln^2 \varepsilon^2 M^2} \over 2} - B \ln \varepsilon^2 M^2
- b_m \right) \tripledelta
$$

\ruline

$$
\Gamma_{n}^b (x_i , \varepsilon ) \to - {g^4 \over \pi^4}
\delta_{12}
\Bigg[ {\dpar \over {\partial x_3^\mu}} \left( {1 \over
x_{14}^2 x_{34}^2} {\lrhup{\partial} \over {\partial x_3^\mu}}
\ln {{x_{13}^2} \over {x_{14}^2}} \int d^4 y {1 \over
{(y-x_1)^2 (y-x_3)^2 (y-x_4)^2}}\right)
$$ $$
+\pi^2 {\dpar \over {\partial x_3^\mu}} \left( {1 \over
{x_{13}^2 x_{14}^2}} {\lrhup{\partial} \over {\partial x_3^\mu}}
{{\ln^2 x_{34}^2 M^2 + 2 \ln x_{34}^2 M^2} \over {x_{34}^2}}
\right)
$$ $$
+ {\pi^4 \over 3} \delta_{13} \dsq {{\ln^3 x_{14}^2 M^2
+ 6 \ln^2 x_{14}^2 M^2 + 12 \ln x_{14}^2 M^2} \over
{x_{14}^2}} \Bigg] + \perm{11}
$$ $$
+ 2g \ln \varepsilon^2 M^2 \Gamma_{h}^b  (x_i , \varepsilon )
- 2 g^2 \left(\ln^2 \varepsilon^2 M^2 + 2\ln \varepsilon^2 M^2
- 2 - 2B \right) \Gamma_f^b  (x_i , \varepsilon )
$$ $$
- 16 \pi^2 g^4 \left( \ln^3 \varepsilon^2 M^2 + 6 \ln^2
\varepsilon^2 M^2  + 6 (1-B) \ln \varepsilon^2 M^2  - 12
- 6B + 3b_n \right) \tripledelta
$$

\ruline

$$
\Gamma_o^b (x_i , \varepsilon ) \to - {g^4 \over 3} \delta_{12}
\delta_{34}
\dsq {{\ln^3 x_{13}^2 M^2 + 3 \ln^2 x_{13}^2 M^2 - 6 \ln
x_{13}^2 M^2} \over {x_{13}^2}} + \perm{2}
$$ $$
+ g \ln \varepsilon^2 M^2 \Gamma_g^b (x_i , \varepsilon ) -
{g^2 \over 2} \left( \ln^2 \varepsilon^2 M^2 + 2\ln \varepsilon^2
M^2 - 2 - 2B \right) \Gamma_f^b (x_i , \varepsilon)
$$ $$
- 4 \pi^2 g^4 \left( \ln^3 \varepsilon^2 M^2 + 3 \ln^2
\varepsilon^2 M^2 - 6 \ln \varepsilon^2 M^2  + b_o \right)
\tripledelta
$$

\ruline

$$
\Gamma_p^b (x_i , \varepsilon ) \to -{{4 g^4} \over \pi^4}
\left( \dsq {{\ln x_{12}^2 M^2} \over {x_{12}^2}} \right) {{x_{12}^2}
\over {x_{13}^2 x_{14}^2 x_{23}^2 x_{24}^2 x_{34}^2}}
$$ $$
- 96 \pi^2 g^4 \left( \zeta (3) \ln \varepsilon^2 M^2 + b_p \right)
\tripledelta $$

\ruline


\sect{5.\ \ Counterterms and Unitarity}

In the previous section, we computed the bare
amplitudes for 2- and 4-point functions. These amplitudes consist of the
renormalized amplitudes of [1] plus terms which diverge as the
ultraviolet cutoff $\varepsilon$ goes to 0. In much the same way as in
any
other regularization method, we now show that the divergent terms can be
cancelled by local counterterms in the Lagrangian. Specifically, the
purely local divergent terms, those involving only delta-functions,
directly
determine counterterms which are added to the Lagrangian. The counterterm
vertices in turn generate non-local graphs which must cancel the
non-local
divergent terms in the bare amplitudes. This last cancellation is
the nontrivial
check of the consistency of the entire procedure and the keystone
of the proof
of unitarity. After it is finished we can all go to the beach.

Of course renormalization conditions are necessarily involved in the
process of determining counterterms, and we implicitly use the
conditions of
[1]. The essential role of these conditions is to fix the finite
parts of the
divergent terms, and the consistent cancellation of these by the
finite parts of
the counterterms below is crucial to unitarity. We remind the
reader that in
the standard form of the renormalization procedure used in this
section, the
coupling $g$ which appears in the Tables is the ``physical''
coupling
within the renormalization scheme of [1].

The simplest illustration of this program is the 1-loop
renormalization of the coupling already discussed in Section 2.
There
we saw how the local divergent term in the bare amplitude (2.14)
of graph
{\ftt f\/} is cancelled by a counterterm in the Lagrangian obtained
from the
coupling shift (2.15).

Let us now study the case of the 2-point function. Its total
bare amplitude is found adding the contributions of graphs {\ftt a\/},
{\ftt c\/}, and {\ftt d\/} from Table 1. The local divergences are
cancelled by the following counterterms :

\ii{\fbx 1.} a wave function renormalization,
$$
-\sq \delta (x) \left( Z_\phi - 1 \right) =  - \sq \delta (x) \left[ -
{g^2 \over {12}} \left( 1 - \ln \varepsilon^2 M^2 \right)  \right.
$$ $$
- \left. {g^3 \over 8} \left( \ln^2 \varepsilon^2 M^2 + \ln
\varepsilon^2 M^2 + {2D_2 \over \pi^2} \right) \right] \, , \eqnn{5.1}
$$

\ii{\fbx 2.} and a perturbative mass counterterm,
$$
\delta m^2 Z_\phi \delta (x)
= \delta (x) {1 \over \varepsilon^2} \left[ {1 \over 3} g^2 - {1 \over
\pi^3} g^2 \left( \pi^2 \ln \varepsilon^2 M^2 - 2 D_1 \right) \right]
\, .  \eqnn{5.2}
$$

Finally, the non-local divergence of graph {\ftt d\/},
$$
3g \ln \varepsilon^2 M^2 \quad \Gamma_c^b (x, \varepsilon) \eqnn{5.3}
$$
cancels when we take into account in the computation of {\ftt c\/},
the one loop modification of the coupling (2.15).
We have, thus, shown how counterterms can
eliminate all divergences, {\it local as well as non-local},
in the two point function. We recover the renormalized amplitude of [1].

The same can be done for the four point function at higher loops. The
local triple-delta
terms are easily absorbed into a redefinition of the vertex. One
includes, then, all the new counterterm vertices, that is coupling,
kinetic and
mass corrections, in the computation of the amplitude and checks
that the non-local divergences cancel. Let us stress the non-triviality
of such cancellation, which is particularly dramatic for the
$\Gamma_f^b$
terms at the three loop level. Here, the sum of the 8 3-loop graphs
$(i-p)$ contains the term
$$
-g^2  \left( {{27} \over 4} \ln^2 \varepsilon^2 M^2 + {{35} \over
6} \ln \varepsilon^2
M^2 - {{35} \over 6} - 6B \right) \Gamma_f^b (x, \varepsilon)
\eqnn{5.4}
$$
and this is neatly cancelled by the counterterm vertices from both 1-
and 2-loop coupling renormalization inserted in graph {\ftt f\/}. In the
end,
we recover the result of [1].

These results can be summarized by adding the counterterms to the
Lagrangian (2.1). The result is the so-called bare Lagrangian,
$$
{\cal L}_{bare} = {1 \over 2} Z_\phi \left( \partial_\mu \phi
\right)^2 + {1 \over 2} Z_\phi
\delta m^2 \phi^2 + {{16 \pi^2} \over {4!}} g Z_g Z_\phi^2 \phi^4,
\eqnn{5.5}
$$
with
$$
Z_g Z_\phi^2  =  1 - {3 \over 2} g \ln \varepsilon^2 M^2 + g^2
\left( {9 \over 4} \ln^2 \varepsilon^2 M^2 + 3 \ln \varepsilon^2
M^2 -3 -3B \right)
$$ $$
+ {1 \over {16}} g^3 \left( -54 \ln^3 \varepsilon^2 M^2 - 178 \ln^2
\varepsilon^2 M^2 \right.
$$ $$
\left. + \left( 216 B - 44 - 96 \zeta (3) \right)
\ln \varepsilon^2 M^2 + c \right) + O \left( g^4 \right)
\, , \eqnn{5.6}
$$

$$
Z_\phi  =  1 - {g^2 \over {12}} \left( 1 - \ln \varepsilon^2 M^2
\right) - {g^3 \over 8}
\left( \ln^2 \varepsilon^2 M^2 + \ln \varepsilon^2 M^2 + {2D_2 \over
\pi^2} \right) + O \left( g^4 \right) \, , \eqnn{5.7}
$$

$$
\delta m^2 Z_\phi = {1 \over \varepsilon^2} \left[ {1 \over 3} g^2 -
{1 \over \pi^2} g^3 \left( \ln \varepsilon^2 M^2 - 2 D_1 \right)
\right] + O \left( g^4 \right) \, . \eqnn{5.8}
$$
We can rewrite {5.5} defining the bare field $\phi_0 =
Z_\phi^{1/2} \phi$, the bare coupling $g_0=gZ_g$ and the bare mass
$m_0^2 = \delta m^2$. Then, the bare Lagrangian becomes
$$
{\cal L}_{\rm bare} = {1 \over 2} \left( \partial_\mu \phi_0 \right)^2 +
{1 \over 2} m_0 \phi_0^2 + {{16 \pi^2} \over {4!}} g_0 \phi_0^4 \, .
\eqnn{5.9}
$$

The constant $c$ in (5.6) is the sum of the various numerical
constants in the $g^4 \delta_{12} \delta_{13} \delta_{14}$ terms of
the 3-loop graphs of Table 2.  Its value would become relevant only if
the calculations of this paper are extended to 4-loop order.


\sect{6.\ \ Renormalization Group Equations}

The renormalizability of {\fit massive\/} $\phi^4$ theory means
that, in any correct regularization procedure with short-distance
cutoff $\varepsilon$ and any renormalization scheme with scale parameter
$M$, renormalized and bare 1PI $n$-point functions are related by
$$
\Gamma^{(n)}_{\rm ren} \left( x_i, g, m, M \right) = Z_\phi^{n \over 2}
\!\left( g_0, M, \varepsilon \right) \Gamma^{(n)}_{\rm bare} \left( x_i,
g_0, m_0, \varepsilon \right) \ . \eqnn{6.1}
$$
The right-hand side has a finite limit as $\varepsilon\to 0$ with
physical
coupling $g$ and mass $m$ held fixed.

The fact that
$$
M {\partial \over \partial M} \Gamma^{(n)}_{\rm bare} \left( x_i,
g_0, m_0, \varepsilon \right) \bigg|_{g_0, m_0, \varepsilon\ {\rm fixed}}
= 0 \eqnn{6.2}
$$
leads directly [4] to the Callan-Symanzik equation
$$
\left[ M {\partial \over \partial M} + \beta (g) {\partial \over
\partial g} - n \gamma (g) - \delta (g) m {\partial \over \partial
m} \right] \Gamma^{(n)}_{\rm ren} \left( x_i, g, m, M\right) = 0
\eqnn{6.3}
$$
with
$$\eqalignno{
\beta (g) &= M {\partial \over \partial M} g \left( g_0,
M, \varepsilon \right) \bigg|_{g_0, \varepsilon\ {\rm fixed}}
&{\fss (6.4)} \cr %
\gamma (g) &= {1 \over 2} M {\partial \over \partial M} \ln Z_\phi
\left( g_0, M, \varepsilon \right) \bigg|_{g_0,\varepsilon\ {\rm fixed}}
&{\fss (6.5)} \cr %
\delta (g) &= - {M \over 2m^2} \ {\partial \over \partial M} m^2
\left( g_0, m_0, M, \varepsilon \right) \ . &{\fss (6.6)} \cr
}
$$
The relations between bare and physical parameters, such as $g_0 =
gZ_g$, must be inverted to compute the functions $\beta(g)$,
$\gamma(g)$, $\delta(g)$.

The cancellation of divergent terms with counterterms in Section 5
establishes that the differential renormalization
procedure is correct through three-loop order for $m=0$, so that (6.1)
holds up to three-loop order in the massless theory.  One can think
of the bare amplitudes as defined via (6.1) by substituting $g$
($g_0$), computed from (5.6)--(5.7), in the renormalized amplitudes
of Ref.~[1] and multiplying by $Z_\phi^{-n/2}$.  However, our
computational
procedure also provides quite directly the explicit form of the bare
amplitudes.  Specifically, the entries in Tables 1 and 2 are the
result of a systematically cutoff computation with Lagrangian coupling
$g$, zero Lagrangian mass and unit normalization of propagators.  Thus
if we simply replace $g\to g_0$ in Table~1, we can interpret the
entries there as $\Gamma^{(n)}_{\rm bare} \left( x_i, g_0, 0,\varepsilon
\right)$, and one can verify directly from the table that the apparent
$M$ dependence of these amplitudes cancels.  To obtain
$\Gamma^{(n)}_{\rm bare} \left( x_i, g_0, m_0,\varepsilon \right)$, one
simply adds the mass insertion (5.2) or (5.8) rewritten in terms of
the bare coupling as
$$
m^2_0 = \delta m^2 = {1 \over \varepsilon^3} \left( {1 \over 3} g^2_0 -
{{2D_1} \over \pi^2} g^3_0 \right) + {\cal O} (g^4_0) \ .\eqnn{6.7}
$$
The sole effect of this is to cancel all quadratically divergent
terms in the entries of the Tables, leaving amplitudes which clearly
satisfy (6.2) because $M{\partial \over \partial M} \delta m^2=0$.

The previous arguments establish the validity of the standard
formulae (6.4)--(6.5) for the renormalization group functions
through three-loop order for $m = 0$.  It is a straightforward
matter to use (5.6)--(5.7), with proper attention to the inversion
of $g_0 (g) \leftrightarrow g (g_0)$, and obtain
$$\eqalignno{
\beta (g) &= 3 g^2 - {17 \over 3} g^3 + \left( 31 + 12\zeta (3)
\right) g^4 + {\cal O} (g^5) &{\fss (6.8)} \cr
\gamma (g) &= {1 \over 12} g^2 - {3 \over 8} g^3 + {\cal O} (g^4)
\ . &{\fss (6.9)} \cr
}
$$
Our calculations have probed only the massless theory, and it is
clear that the $m{\partial \over \partial m}$ term in (6.3) vanishes
as $m \to 0$ because there are no infrared divergences.  Therefore,
we do not discuss $\delta (g)$ here because it requires information
about $m \not = 0$.  The results for $\beta (g)$, $\gamma (g)$ found
here in the standard framework of the renormalization group equations
agree with those of the ``experimental'' approach taken in Ref.~[1].
This provides another check that the differential renormalization
procedure is correct.


\sect{7.\ \ An Alternate Cutoff Method}

The cutoff method used in Sections 1--4, which is based on the damped
propagator (1.2), is actually the second method we have applied to
this problem.  In our first approach, which was described briefly in
the first article of Ref.~[1], regularization of bare amplitudes was
achieved by the exclusion of small balls of radius $\varepsilon$ about
short distance singularities.  Integrals involving such cutoff bare
amplitudes then converge, and the singular contributions as
$\varepsilon \to 0$ are quite clearly related  to the surface terms
dropped in the partial integration rule of differential
renormalization.

The systematic rules used in this regularization method were the
following:

\i{1)}{\fss Each propagator connecting vertices $x$ and $y$ of a diagram
is replaced by the cutoff propagator
$$
\Delta (x-y) = {1 \over {4 \pi^2}} {1 \over {(x-y)^2}} \ \rightarrow
\ \Delta (x-y, \varepsilon) = {1 \over {4 \pi^2}} {{\Theta \left( | x-y
| - \varepsilon \right)} \over {(x-y)^2}} \eqnn{7.1}
$$
where $\Theta (z)$ is the step function.}

\i{2)}{\fss For each pair of (internal or external) vertices $x_i, x_j$
not connected by a propagator, the bare amplitude is multiplied by an
additional cutoff factor $\Theta \left( | x_i - x_j | - \varepsilon
\right)$.}

Calculations using this approach were generally simpler than in the
current method because the differential identities of the Appendix of
[1] could be used directly.  Further, after partial integration, one
finds $\delta \left( | x_i - x_j | - \varepsilon \right)$ terms when
derivatives act on the step function cutoff factors, and these
effectively reproduce the surface terms which are the crucial
issue.  Complete results through 3-loop order were obtained, and we
found that the cutoff amplitudes for each graph could be expressed as
the renormalized amplitudes of [1], plus singular terms which could
be consistently compensated by counterterms in the Lagrangian.  The
renormalization group functions $\beta (g)$ and $\gamma (g)$ were
calculated from the cutoff dependence of $Z_g$ and $Z_\phi$, as in (6.4)
and (6.5), again with results identical to those of [1].

Despite the successful result and relative
ease of calculation, we now believe that this
method does not support the conclusion that the renormalized amplitudes
of [1] satisfy perturbative unitarity.
To discuss this we first note that the cutoff propagator (7.1) has
Fourier transform
$$
\Delta (p, \varepsilon) = {1 \over p^2} J_0 (p \varepsilon) \eqnn{7.2}
$$
where $J_0 (z)$ is the Bessel function which is analytic in its
argument.  Since the only singularity of $\Delta (p, \varepsilon)$ is
the standard $1/p^2$ pole, the method would give a plausible argument
for unitarity provided that calculations could be done using only the
propagator cutoff of Rule 1) above.

In principle, the propagator cutoff is sufficient to make all required
integrals converge, but  it was technically too difficult to do many
integrals in this way.  Instead we adopted the procedure of performing
subintegrals using Rule 1), but then substituting the limiting form of
this result before studying further singularities of a graph which
were cutoff by factors from Rule 2).
These Rule 2) cutoff factors
cannot be described as a modification of the Lagrangian which is
Hermitean below some cutoff energy scale, and this raises more
questions about unitarity.

In view of the above, one may wonder whether the result of the
consistent counterterms mechanism found in this method was an accident
or whether it encapsulates some truth.  We think that the latter is
correct, because our procedures, albeit somewhat sloppy, were used
consistently.  Subgraphs of a given graph were handled by the same
steps as in their initial appearance in lower order.

Very recently, it has been shown that $x$-space dimensional
regularization can be combined with differential identities so as to
reproduce several of the lower order amplitudes of [1] plus local
counterterms [5]. In higher order, this method could lead to a useful
relation between the amplitudes of the differential renormalization
and dimensional regularization procedures.


\sect{8.\ \ Concluding Remarks}

We believe that the calculations of Secs.~2 -- 4 have fulfilled their
intended goals.  Namely, a systematic real space cutoff method for
$\phi^4$ theory has been used to show that through 3-loop order, the
bare 2- and 4-point correlation functions can be expressed as the sum
of the renormalized amplitudes of [1] plus a combination of singular
and finite terms.  This combination can be compensated by adding the
traditional counterterms to the Lagrangian.  Indirectly this
demonstrates that the major heuristic rule of the differential
renormalization procedure, namely formal partial integration, is
consistent.  Since the cutoff method is based on a damped propagator
whose Fourier transform (1.4) consists of the usual pole plus a cut
whose effects vanish as $\varepsilon^2$, our results also imply that
differential renormalization obeys perturbative unitarity.  Finally we
have shown that the same renormalization group functions $\beta (g)$ and
$\gamma (g)$ are obtained in the cutoff theory and in the method of
[1], and this is an additional consistency check.

An important subsidiary purpose of our work was to convince skeptics
that overlap divergences are correctly treated in differential
renormalization. The results above do demonstrate this since all
non-local divergences are exactly cancelled by the local counterterms
added to the lagrangian. However, it
may be useful to restate and amplify upon the common belief that overlap
divergences are not a problem in real space calculations, because
subdivergent regions remain distinct and can be regulated before the
overall divergence of a graph is studied.  Let us illustrate this in
the case of the most conspicuous overlap graph in our work, the 3-loop
cateye graph {\ftt o\/}.
We note that the treatment of this graph in [1] started with the
expression
$$ \eqalign{
\Gamma_o (x_i) &= - {{4g^4} \over \pi^4} \delta_{12}
\delta_{34} f_o (x_{13}) + \perm{2} \cr
f_o (x) &= - {1 \over 4} \int {{d^4 u \, d^4 v} \over {u^2 v^2 (x-u)^2
(x-v)^2}} \dsq {{\ln (u-v)^2 M^2} \over {(u-v)^2}}\ . \cr
} \eqnn{8.1}
$$
This expression can also be obtained from (4.45) by setting
$\varepsilon = 0$ and using (1.1) to regulate the central bubble
subgraph.

There are three subdivergent regions in this graph; namely the
4-dimensional region $u \sim v$, and the eight dimensional regions $u
\sim v \sim 0$ and $u \sim v \sim x$.  The first of these is already
regulated by the use of (1.1) which is an identity for $u - v \not =
0$, and is defined at the singularity by the partial integration rule.
This means that the divergent 1-bubble subgraph is treated exactly in
the same way as the renormalized amplitude for graph {\ftt f\/}.  After
partial integration of $\sq_u$ in (8.1) one finds [1] two integrals
with $\delta (u)$ and $\delta (x-u)$ factors together with a cross
term, as one can see from (4.46) at $\varepsilon = 0$.  The cross term
is complicated but contains no subdivergences.  Indeed the $\delta (u)$
and $\delta (x-u)$ factors give a partial localization of the
8-dimensional singular regions, and it is not difficult to see that the
integrand in these regions is treated in [1] exactly as the
product of the renormalized amplitude of the ice cream cone subgraph
(specifically, the first term in the entry for $\Gamma_h^b$ in Table 2
with correct designation of variables) times the remaining
non-singular propagator factors.  The internal integrals $d^4 u \, d^4
v$ are then performed leading to the explicit form of $f_o (x)$
containing $( \ln x^2 M^2)^n / x^4$ overall singularities which are
easily regulated.  Independent of a systematic cutoff
procedure, our confidence that overlap divergences are properly
treated in differential renormalization is based on the property that
the amplitude in subdivergent regions is regulated in the same way
as the appropriate subgraph.

\vskip 2pc
\ni{\flbx Acknowledgements} \nobreak \vskip 3pt

This work is supported in part by funds
provided by the U. S. Department of Energy (D.O.E.) under contract
\#DE-AC02-76ER03069, the National Science Foundation under grant
\#87-08447, and CICYT. RMT acknowledges a Fleming fellowship from the
Ministerio de Educaci\'on y Ciencia jointly with the British Council.
XVC acknowledges an FPI grant from the Ministerio de Educaci\'on y
Ciencia.

We would like to thank D.~Espriu, E.W.N.~Glover, P.E.~Haagensen,
C.~Kounnas, J.~Soto
and R.~Tarrach for useful discussions. We would like to thank very
especially J.I.~Latorre for his help and support during the achievement
of this work.

\sect{References}

\ent{[1]}{D.Z.~Freedman, K.~Johnson, J.I.~Latorre,
Nucl.~Phys. {\fbx B371,\/} 329 (1992).}

\ent{[2]}{D.Z.~Freedman, ``Differential regularization and
renormalization: recent progress".  In {\fsl Proceedings of the
Stony Brook Conference on Strings and Symmetries,\/} Spring, 1991.}

\ent{}{P.E.~Haagensen, Mod.~Phys.~Lett.~{\fbx A7,\/} 893 (1992).}

\ent{}{D.Z.~Freedman, G.~Grignani, K.~Johnson, N.~Rius, ``Conformal
symmetry and differential regularization of the 3-gluon vertex", MIT
preprint, CTP \#1991, to appear in Annals of Physics.}

\ent{}{P.E.~Haagensen, J.I.~Latorre, ``Differential renormalization of
massive quantum field theories", Universitat de Barcelona preprint,
UB-ECM-PF 92/5, to appear in Physics Letters {\fbx B \/}.}

\ent{}{R.~Mu\~noz-Tapia, ``A new approach to lower
dimensional gauge theories", University of Durham preprint, DTP/92/34.}

\ent{[3]}{J.~Polchinski, Nucl.~Phys.~{\fbx B231,\/} 269 (1984).}

\ent{[4]}{C.G.~Callan, Phys.~Rev.~{\fbx D2,\/} 1541 (1970).}

\ent{}{K.~Symanzik, Comm.~Math.~Phys.~{\fbx 18,\/} 227 (1970).}

\ent{}{D.J.~Gross, in Methods in Field Theory, {\fsl Proceedings of
the Les Houches Summer School 1975,\/} edited by R.~Balian and J.~Zinn
Justin, North Holland, 1976.}

\ent{}{C.~Itzykson, J.B.~Zuber, Quantum Field Theory, McGraw-Hill
(1980).}

\ent{[5]}{G.~Dunne, N.~Rius, in preparation.}

\sect{Appendix}

We include in this Appendix some technical results.  Most of them have
been used in the calculations of several of the graphs discussed in
Sections 2 -- 4 of the main text.

\insect{1.}
The convolution integral
$$
I_\varepsilon (x) = \int d^4 u {1 \over {[u^2 + \varepsilon^2]^2 \left[
(x-u)^2 + \varepsilon^2 \right]^2}} \eqnn{A.1}
$$
is required to evaluate graphs {\ftt d\/}, {\ftt g\/}, {\ftt i\/}, and
{\ftt j\/}.  The integral can be done using the fact (see (2.12)) that
the Fourier transform of $1/(u^2 + \varepsilon^2)^2$ is $2 \pi^2 K_0
(p \varepsilon)$.  We can write
$$ \eqalign{
I_\varepsilon (x) &= \int {{d^4 p} \over {(2 \pi )^4}} e^{- i p \cdot
x} 4 \pi^4 K_0^2 (p \varepsilon) \cr
{} &= - {1 \over 4} \sq \int d^4 p {e^{- i p \cdot x} \over p^2} K_0^2
(p \varepsilon ) \cr
{} &= - \pi^2 \sq \left\{ {1 \over x} \int_0^\infty dp J_1 (px) K_0^2
(p \varepsilon ) \right\} \cr
{} &= - \pi^2 \sq \left\{ {{\left[ {\rm arctanh} \left( \sqrt {1 + {{4
\varepsilon^2} \over x^2}} \right) \right]^2} \over x^2} \right\} \cr
{} &= - {\pi^2 \over 4} \sq {{\ln^2 \left[ \left( x^2 + 2
\varepsilon^2 - x \sqrt {x^2 + 4 \varepsilon^2} \right) / 2
\varepsilon^2 \right] } \over x^2} \ . \cr
} \eqnn{A.2}
$$
This is an exact result.  As $\varepsilon \rightarrow 0$ we obtain
$$
I_\varepsilon (x) \rightarrow - {\pi^2 \over 4} \dsq {{\ln^2 (x^2 /
\varepsilon^2 ) } \over x^2} \ . \eqnn{A.3}
$$
If we introduce the mass scale $M$, this can be rewritten as
$$
I_\varepsilon (x) \rightarrow - {\pi^2 \over 4} \dsq {{\ln^2 x^2 M^2}
\over x^2} + {\pi^2 \over 2} \ln \varepsilon^2 M^2 \dsq {{\ln x^2 M^2}
\over x^2} + \pi^4 \delta (x) \ln^2 \varepsilon^2 M^2 \ . \eqnn{A.4}
$$

This is the same result we would have obtained if the limiting form of
the regulated bubble
$$
{1 \over {[u^2 + \varepsilon^2]^2}} \rightarrow - {1 \over 4} \dsq
{{\ln x^2 M^2} \over x^2} - \pi^2 \ln \varepsilon^2 M^2 \delta (x) \ ,
\eqnn{A.5}
$$
were inserted in (A.1), and the integral computed using formal partial
integration as in [1].  In other words, it is justified in this case
to take the limit in the integrand.  The reason for this appears to
lie in (2.8) in which the cutoff bubble amplitude is expressed as
$\sq$ of a function which has a soft singularity as $\varepsilon
\rightarrow 0$.  The $\sq$ operator can be transferred to the external
variables in (A.1), leaving a function which is smooth enough that the
$\varepsilon \rightarrow 0$ limit can be taken in the integrand.  This
is sufficient for the evaluation of graphs {\ftt g\/} and {\ftt i\/}.
However, in graphs {\ftt d\/} and {\ftt j\/}, where other singular
factors multiply $I_\varepsilon (x)$, the more accurate form (A.2) is
required to study the limit of the bare amplitudes.

\insect{2.}
Representations of the distributions $\delta (x)$ and $\sq \delta (x)$
appear throughout our work.  The simplest example is the basic
equation for the cutoff propagator,
$$
\sq \Delta (x, \varepsilon) = {1 \over {4 \pi^2}} \sq {1 \over {(x^2 +
\varepsilon^2)}} = - {{2 \varepsilon^2} \over {\pi^2 (x^2 +
\varepsilon^2)^3}} \ . \eqnn{A.6}
$$
Of course we expect the limiting relation
$$
{{2 \varepsilon^2} \over {\pi^2 (x^2 + \varepsilon^2)^3}} \rightarrow
\delta (x) \ . \eqnn{A.7}
$$
To prove this we integrate this candidate $\delta$-function with a
smooth test function $f(x)$ which is damped at long distances.  We
write
$$ \eqalign{
\int d^4 x f(x) {{2 \varepsilon^2} \over {\pi^2 (x^2 + \varepsilon^2
)^2}} &= {2 \over \pi^2} \int d^4 y \, f(\varepsilon y) \, {1 \over
{(y^2 + 1)^3}} \cr
{} &= {2 \over \pi^2} f(0) \int {{d^4 y} \over {(y^2 + 1)^3}} +
{2 \over \pi^2} \int
d^4 y {{f (\varepsilon y) - f(0)} \over {(y^2 + 1)^3}} \ . \cr
}
$$
It follows from Taylor's theorem that $f(\varepsilon y) - f(0) \sim
{\cal O} (\varepsilon)$ as $\varepsilon \rightarrow 0$ so the limit of
the second infrared convergent integral vanishes.  The first integral
is easy to evaluate,
$$ \eqalign{
\int {{d^4 y} \over {(y^2 + 1)^3}} &= \int d \hat y \int_0^\infty
{{y^3 dy} \over {(y^2 + 1)^3}} \cr
{} &= 2 \pi^2 \int_0^\infty {{y^3 dy} \over {(y^2 + 1)^3}} = {\pi^2
\over 2} \ . \cr
}
$$

Other representations of distributions are
$$
{{3 \varepsilon^2} \over {(x^2 + \varepsilon^2)^4}} \rightarrow {\pi^2
\over {2 \varepsilon^2}} \delta (x) + {\pi^2 \over 8} \sq \delta (x)
\eqnn{A.8a}
$$ $$
{{\varepsilon^2 \ln (x^2 + \varepsilon^2) M^2} \over {x^2 (x^2 +
\varepsilon^2)^2}} \rightarrow \pi^2 (\ln \varepsilon^2 M^2 + 1)
\delta (x)  \ . \eqnn{A.8b}
$$
When $\sq \delta (x)$ is involved, it is necessary to expand the test
function in a Taylor series through second order in order to extract
the limiting form.  Similar relations which hold as $\varepsilon
\rightarrow 0$ are
$$
{1 \over {(x^2 + \varepsilon^2)^2}} = {1 \over {x^2 (x^2 +
\varepsilon^2)}} - \pi^2 \delta (x) \eqnn{A.9a}
$$ $$
{{\ln (x^2 + \varepsilon^2)/\varepsilon^2} \over {(x^2 +
\varepsilon^2)^2}} = {{\ln (x^2 + \varepsilon^2) / \varepsilon^2}
\over {x^2 (x^2 + \varepsilon^2)}} - \pi^2 \delta (x)
\eqnn{A.9b}
$$ $$
{{\ln^n (x^2 + \varepsilon^2) / \varepsilon^2} \over {(x^2 +
\varepsilon^2)^2}} = {{\ln^n (x^2 + \varepsilon^2)/\varepsilon^2}
\over {x^2 (x^2 + \varepsilon^2)}} - \pi^2 n ! \delta (x) \ .
\eqnn{A.9c} $$

\insect{3.}
Differential Regularization identities for cutoff singular functions
are useful throughout our calculations.  These include the following
identities,
$$
{1 \over {(x^2 + \varepsilon^2)^2}} = - {1 \over 4} \sq {{\ln (x^2 +
\varepsilon^2)/\varepsilon^2} \over x^2} \eqnn{A.10a}
$$ $$
{1 \over {(x^2 + \varepsilon^2)^3}} = {1 \over {32}} \sq \sq {{\ln
(x^2 + \varepsilon^2) / \varepsilon^2} \over x^2} + {{3 \varepsilon^2}
\over {(x^2 + \varepsilon^2)^4}} \eqnn{A.10b}
$$ $$
{{\ln (x^2 + \varepsilon^2) / \varepsilon^2} \over {[x^2 +
\varepsilon^2]^2}} = - {1 \over 8} \sq {{\ln^2 (x^2 + \varepsilon^2) /
\varepsilon^2 + 2 \ln (x^2 + \varepsilon^2) / \varepsilon^2} \over
x^2} \ . \eqnn{A.10c}
$$

As $\varepsilon \rightarrow 0$, we also have
$$
{{\ln (x^2 + \varepsilon^2) M^2} \over {x^2 (x^2 + \varepsilon^2)}}
\rightarrow - {1 \over 8} \sq {{\ln^2 (x^2 + \varepsilon^2) M^2 + 2
\ln (x^2 + \varepsilon^2) M^2} \over x^2} - {\pi^2 \over 2} \left(
\ln^2 \varepsilon^2 M^2 - 2 \right) \delta (x) \eqnn{A.11a}
$$ $$
{{\ln^2 (x^2 + \varepsilon^2)M^2} \over {x^2 (x^2 + \varepsilon^2)}}
\rightarrow
- {1 \over {12}} \sq {{\ln^3 (x^2 + \varepsilon^2) M^2 + 3 \ln^2 (x^2 +
\varepsilon^2) M^2 + 6 \ln (x^2 + \varepsilon^2) M^2} \over x^2}
$$ $$
- \pi^2 \left( {1 \over 3} \ln^3 \varepsilon^2 M^2 - {1 \over 2} \right)
\delta (x)  \ . \eqnn{A.11b}
$$

\insect{4.}
Triangular structures have the schematic form,
$$
\big[ {\rm a\ representation\ of\ }\delta (x-y) \big] \times \left[
{{{\rm smooth\ } f(y)} \over {(y^2 + \varepsilon^2) {\rm \ or\ } y^2}}
\right] \times \left[ {{{\rm smooth\ } g(x)} \over {(x^2 +
\varepsilon^2) {\rm \ or\ } x^2}} \right]
$$
Such structures first appear in the 2-loop graph {\ftt h\/}, and we
also encounter them in many 3-loop graphs.  The basic strategy to
obtain their limiting form, which involves integration with a test
function $f(x,y)$ of two variables was discussed in connection with
graph {\ftt h\/}.  Using this strategy one obtains the limiting forms
$$
\sq \left[ {1 \over {(x-y)^2 + \varepsilon^2}} \right] {1 \over {x^2 +
\varepsilon^2}} {1 \over y^2} \rightarrow - 4 \pi^2 \delta (x-y) {1
\over {y^2
(y^2 + \varepsilon^2)}} + 4 \pi^4 \delta (x) \delta (y) \eqnn{A.12a}
$$ $$
\sq \left[ {1 \over {(x-y)^2 + \varepsilon^2}} \right] {1 \over {x^2 +
\varepsilon^2}} {{\ln (y^2 + \varepsilon^2 ) M^2} \over y^2} \rightarrow
-
4 \pi^2 \delta (x-y) {{\ln (y^2 + \varepsilon^2) M^2} \over {y^2 (y^2 +
\varepsilon^2)}} + 4 \pi^4 ( \ln \varepsilon^2 M^2 - B) \delta (x)
\delta (y) \eqnn{A.12b}
$$ $$
\sq \left[ {1 \over {(x-y)^2 + \varepsilon^2}} \right] {1 \over {x^2 +
\varepsilon^2}} {{\ln^n (y^2 + \varepsilon^2 ) / \varepsilon^2} \over
y^2} \rightarrow - 4 \pi^2 \delta (x-y) {{\ln^n (y^2 + \varepsilon^2) /
\varepsilon^2} \over {y^2 (y^2 + \varepsilon^2 )}} - 4 \pi^4 B_n
\delta (x) \delta (y) \eqnn{A.12c}
$$ $$
\sq \left[ {1 \over {(x-y)^2 + \varepsilon^2}} \right] {{\ln (x^2 +
\varepsilon^2 ) / \varepsilon^2} \over {x^2 + \varepsilon^2}} {1 \over
y^2} \rightarrow - 4 \pi^2 \delta (x-y) {{\ln (y^2 + \varepsilon^2 ) /
\varepsilon^2} \over {y^2 (y^2 + \varepsilon^2)}} + 4 \pi^4 \delta (x)
\delta (y) \ . \eqnn{A.12d}
$$

The coefficient of $\delta (x) \delta (y)$ in (A.12b) is simply the
double integral of the last term in (3.16).  This leads to the
following expression for $B$,
$$
B={1 \over {4 \pi^4}} \int d^4x\, d^4y\, \sq \left[ {1 \over
{(x-y)^2 + 1}} \right]  {1 \over {x^2(x^2
+ 1)}} {{\ln (y^2 +1)} \over y^2}
$$ $$
=-{1 \over \pi^4} \int d^4 x \, d^4 y \, {1 \over \left( x-y \right)^2 +
1} \, \, \, {1 \over x^2 \left( x^2+ 1 \right)}
\left[ {1 \over y^2 + 1} \right]^2
$$ $$
= {{2 \pi^2} \over {9}} - {1 \over 3} \psi^\pr \left( {1 \over 3}
\right) \approx - 1.171953 \ . \eqnn{A.13}
$$
One can go from the first to the second line of (A.13) by integrating
$\sq$ by parts back onto the log. Finally, this
last integral is computed using the standard Feynman parametrisation.
Similar expressions can be found for the numerical constants $B_n$.
For example, the coefficients of the triple delta terms can be read off
by integrating the difference of both sides of the
equalities (A.12).

\vfill \end